\documentclass[aps,pre,twocolumn,superscriptaddress,nofootinbib,showkeys]{revtex4-2}

\usepackage[utf8]{inputenc}
\usepackage[T1]{fontenc}
\usepackage{graphicx}
\usepackage{amsmath} 
\usepackage{amssymb} 
\usepackage{rotating}
\usepackage{makecell}
\usepackage{array,multirow}
\usepackage{bbm}
\usepackage[normalem]{ulem}
\usepackage{epsfig}
\usepackage{dcolumn}
\usepackage{bm}
\usepackage{xcolor}
\usepackage{hyperref}
\usepackage{orcidlink}

\newcommand{\req}[1]{Eq.~(\ref{#1})}

\newcommand{\fig}[1]{Fig.~\ref{#1}}

\newcommand{\cut}[1]{{}}

\begin{document}
\renewcommand{\topfraction}{0.95}
\renewcommand{\dbltopfraction}{0.95}
\renewcommand{\textfraction}{0.02}
\renewcommand{\floatpagefraction}{0.70}
\renewcommand{\dblfloatpagefraction}{0.70}

\title{Analysis of Developing Cortical Neuronal Networks Using Visual Informatics}

\author{Ho Fai Po\,\texorpdfstring{\orcidlink{0000-0002-3056-4064}}{}}
\email{h.po@aston.ac.uk}
\affiliation{College of Engineering and Physical Sciences, Aston University, Birmingham B4 7ET, United Kingdom}

\author{Akke Mats Houben\,\texorpdfstring{\orcidlink{0000-0002-8215-7882}}{}}
\affiliation{Departament de Física de la Matèria Condensada, Universitat de Barcelona, Barcelona, Spain}
\affiliation{Universitat de Barcelona Institute of Complex Systems (UBICS), 08028 Barcelona, Spain}

\author{Anna-Christina Haeb\,\texorpdfstring{\orcidlink{0009-0008-0194-1036}}{}}
\affiliation{Departament de Física de la Matèria Condensada, Universitat de Barcelona, Barcelona, Spain}
\affiliation{Universitat de Barcelona Institute of Complex Systems (UBICS), 08028 Barcelona, Spain}
\affiliation{Laboratory of Neural Stem Cells and Brain Damage, Department of Biomedical Sciences, Institute of Neurosciences, University of Barcelona, 08036 Barcelona, Spain}
\affiliation{Institut d’Investigacions Biomèdiques August Pi i Sunyer (IDIBAPS), 08036 Barcelona, Spain}

\author{Yordan P. Raykov\,\texorpdfstring{\orcidlink{0000-0003-0753-717X}}{}}
\affiliation{School of Mathematical Sciences, University of Nottingham, Nottingham, UK}

\author{Daniel Tornero\,\texorpdfstring{\orcidlink{0000-0002-4812-4091}}{}}
\affiliation{Laboratory of Neural Stem Cells and Brain Damage, Department of Biomedical Sciences, Institute of Neurosciences, University of Barcelona, 08036 Barcelona, Spain}
\affiliation{Institut d’Investigacions Biomèdiques August Pi i Sunyer (IDIBAPS), 08036 Barcelona, Spain}
\affiliation{Centro de Investigación Biomédica en Red sobre Enfermedades Neurodegenerativas (CIBERNED), 28029 Madrid, Spain}

\author{Jordi Soriano\,\texorpdfstring{\orcidlink{0000-0003-2676-815X}}{}}
\affiliation{Departament de Física de la Matèria Condensada, Universitat de Barcelona, Barcelona, Spain}
\affiliation{Universitat de Barcelona Institute of Complex Systems (UBICS), 08028 Barcelona, Spain}

\author{David Saad\,\texorpdfstring{\orcidlink{0000-0001-9821-2623}}{}}
\affiliation{College of Engineering and Physical Sciences, Aston University, Birmingham B4 7ET, United Kingdom}

\date{\today}

\begin{abstract}
Understanding how neuronal population activity changes during development and after stimulation is essential for studying neuronal network dynamics. This work examines how visual informatics can summarize high-dimensional spiking activity while retaining information that is biologically interpretable. We develop a framework based on Minimum-Distortion Embedding (MDE), and compare it with Principal Component Analysis (PCA) and t-distributed Stochastic Neighbor Embedding (t-SNE). In addition to evaluating the embeddings by visual separation, we quantify whether they preserve the cosine-shape radius within each condition and the pairwise distances between condition centroids. Our \emph{in silico} experiments show that MDE with a cosine metric captures the trajectory of simulated network maturation and preserves the contraction of the activity cloud as connectivity increases. Complementary \emph{in vitro} experiments on human cortical cultures show a coherent developmental trajectory from Day In VITRO 23 (DIV23) to DIV64. We also study weak and strong stimulation in simulation, and long-term potentiation stimulation in primary cortical cultures. In the stimulation experiments, MDE separates activity phases more clearly than PCA and preserves transient changes in within-phase variability that are missed by PCA. These results show that metric selection is central to dimensionality reduction of neuronal data. In particular, cosine distance between population activity vectors provides embeddings that better reflect changes in population activity patterns than Euclidean distance. The proposed framework provides a quantitative way to visualize network development and stimulation-induced changes in neuronal activity.
\end{abstract}




\maketitle

\section{Introduction}
The intricate workings of the brain remain one of the most challenging frontiers in neuroscience. Understanding the changing dynamics of cortical neuronal activity, both over time and in response to stimulation, is essential for unraveling the complexities of neuronal networks and their roles in cognitive functions. Recent advancements in \textit{in-vitro} recording techniques, such as microelectrode arrays (MEAs)~\cite{maccione2010experimental} and calcium-fluorescence imaging~\cite{grienberger2012imaging, kim2022fluorescence}, have enabled the collection of extensive spiking data from neuronal cultures over prolonged periods, often spanning weeks. This wealth of data provides a unique opportunity to explore spiking patterns and connectivity in neuronal networks with unprecedented detail.

Analyzing long-term spiking data through visual informatics, where high-dimensional data are projected onto low-dimensional spaces, highlights the most informative features of the data in an intuitive manner~\cite{Raykov2022}. This approach offers key insights into neuronal network dynamics by revealing spiking patterns over time and illustrating changes in network interactions and connectivity. Such analyses deepen our understanding of how neurons influence each other, elucidating the balance between stability and plasticity that underpins learning and memory. Indeed, \cite{cunningham2014dimensionality} provides a seminal review on how visual informatics techniques enhance exploratory analyses of large datasets, enable single-trial hypotheses to detect significant shifts in neuronal firing, and facilitate population response structure testing.

At the same time, a low-dimensional plot is only useful if the geometry of the embedding can be related back to the original neuronal activity. In this work, we therefore ask whether an embedding preserves both the relative displacement between experimental conditions and the spread of activity within each condition. The first quantity describes how far different developmental or stimulation states are from one another. The second quantity describes whether the population activity within one state becomes more concentrated or more variable.

These methods are also critical for investigating neurological disorders, as they reveal shifts in neuronal activity that can guide disease modeling and therapeutic development~\cite{sotirakis2023identification, iosa2022principal}. By clarifying the mechanisms driving neuronal behavior over time, such research informs our understanding of pathophysiological changes in diverse conditions. These findings may, in turn, inspire advances in neurotechnologies, including brain-machine interfaces and rehabilitation strategies.

Beyond fundamental neuroscience, insights into neuronal activity are increasingly important in the emerging field of neurocortical computation, which seeks to harness human neuronal networks (hNNs) for computing. This initiative aims to address growing concerns about the energy demands and computational limitations of contemporary machine learning (ML) and artificial intelligence (AI) systems~\cite{strubell2019energy}. Notable successes, such as using cortical brain organoids for non-linear curve prediction~\cite{cai2023brain} and for decision-making in simulated gaming environments~\cite{kagan2022vitro}, underscore the potential of this approach. Plasticity and learning in these neuron-based computing devices~\cite{daddinounou2024bi, pedretti2017memristive, li2023short} are generally believed to arise through appropriate stimulation, reshaping network topology and synaptic strengths to perform specific tasks. To evaluate whether a system can successfully execute these tasks, its activity can be analyzed through visual informatics techniques.

Neuronal activity is inherently high-dimensional as it is generated by the interaction of large numbers of interconnected components (neurons), presenting significant challenges in analysis and interpretation~\cite{badre2021dimensionality, mccready2022multielectrode, chung2021neural}. The vast amount of data generated from neuronal recordings can obscure meaningful insights, necessitating advanced analytical techniques such as visual informatics for dimensionality reduction and an intuitive yet informative presentation. Among the various methods employed within the field of visual informatics, dimensionality reduction techniques such as Principal Component Analysis (PCA)~\cite{pearson1901liii, pang2016dimensionality, shinn2023phantom, kobak2016demixed} and t-distributed Stochastic Neighbor Embedding (t-SNE)~\cite{van2008visualizing, zhou2020visualization, hu2020t, kobak2019art} have gained prominence: PCA for its conceptual and computational simplicity and t-SNE perhaps for its ability to identify and conserve clustering information from the high-dimensional space into a lower one. However, PCA assumes a representation in \textit{orthogonal} eigenspaces, which may be overly simplistic for highly interacting neuronal correlations. Conversely, t-SNE excels at preserving \textit{local} (i.e. points close to each other in the original data space) similarity between data points, but points far apart in the original data space (i.e. \textit{global structure}) are represented at an arbitrary distance with respect to each other in the lower dimensional projection. We demonstrate how both of these assumptions might be too unrealistic to facilitate understanding the overarching dynamics of neuronal networks over extensive periods. 

Given these limitations, there is a clear need for dimensionality reduction methods that balance the preservation of both local and global structures, facilitating the retention of information across time and stimuli. To address this gap, we employ \textit{Minimum-Distortion Embedding} (MDE)~\cite{badoiu2005low, fomin2011exact, agrawal2021minimum}, which we further adapt for exploratory analysis of neuronal activity data. We treat each time window as a population activity vector and compare these vectors using cosine distance, so that the embedding emphasizes the shape of the activity pattern across neurons or electrodes. We then evaluate the embedding with two complementary diagnostics. The cosine-shape radius measures the spread of activity patterns within a condition, while pairwise condition-centroid distances measure the displacement between developmental or stimulation states. We compare MDE with traditional dimensionality reduction techniques, such as PCA and t-SNE, highlighting its advantages in capturing the intricate relationships within neuronal activity data. Our evaluation is based on three distinct experimental setups: (i) \textit{in-vitro} experiments observing the development of human-induced pluripotent stem cells (iPSC) derived cortical neuronal networks over a month, (ii) \textit{in-vitro} experiments examining primary cortical neuronal networks before and after long-term potentiation (LTP) stimulation \cite{Chiappalone08, zhu2015different, escobar2024long} within a single day, and (iii) \textit{in-silico} experiments designed to study network development and stimulation effects. Through these comparisons, we show that MDE with a cosine metric preserves both the trajectory of the condition centroids and the changes in within-condition activity variability.

\section{Data Preprocessing}
Consider a system consisting of $N$ neurons within an interactive network. We denote a binary variable $S^r_i(t)$, where $S^r_i(t) = 1$ indicates that neuron $i$ is spiking and $S^r_i(t) = 0$ indicates that it is silent at time $t \in [0, P]$. Here, $r = 1, \ldots, R$ denotes the experimental conditions (such as age or stimulation) under which the network activity is recorded, and $P$ is the total duration of the observation period. The firing pattern of the network at time $t$ is then represented as $\boldsymbol{S}^r(t) = \{ S^r_i(t) \}_{i=1}^N$. Noted that neuron spiking activity is considered as a stochastic process with random fluctuations arising from factors such as ionic channel variability, synaptic noise, and measurement imprecision. Moreover, as each $S^r_i(t)$ can take on two values, the complete firing pattern $\boldsymbol{S}^r(t)$ can exhibit $2^N$ possible states, which can become increasingly noisy as $N$ increases.

To mitigate this noise, we compute the average firing rate over a short time window rather than relying on the binary firing patterns $\boldsymbol{S}^r(t)$. Specifically, for a given window size $\delta$, we define a real-valued variable $u^r_i(\tau) = \left<S^r_i(t)\right>_{t \in (\tau\delta - \delta, \tau \delta)}$, which represents the average firing rate of neuron $i$ during the discrete time window $\tau$, for $i = 1, \ldots, N$ (average is represented by the angled brackets). The collective firing rate across all neurons in condition $r$ is then denoted as ${U}^r(\tau) = \{ u^r_i(\tau) \}_{i=1}^N$, resulting in ${U}^r(\tau) \in \mathbb{R}^N$. This transformation not only enhances the signal-to-noise ratio, but also reduces the dimensionality of the data, facilitating a clearer analysis of neuronal dynamics. The window size $\delta$ is chosen according to the level of data noise.

This data preprocessing step results in a data matrix $\boldsymbol{U} = \left( {U}^r\left(\tau\right) \right)_{r, \tau}$ of size $R\mathcal{T} \times N$, where $\mathcal{T} = \lceil \frac{P}{\delta} \rceil$ is the total number of time windows. Each entry $u^r_i(\tau)$ represents the average firing rate of neuron $i$ during the time window $\left( \tau\delta - \delta, \tau \delta\right)$ under condition $r$. By structuring the data in this manner, we can better capture the temporal dynamics of neuronal activity and facilitate subsequent analyses, such as dimensionality reduction and pattern recognition, which are crucial for understanding the underlying neural processes.

In this analysis, each row of $\boldsymbol{U}$ is treated as a population activity vector. For two population activity vectors $\mathbf{a}$ and $\mathbf{b}$, we use the cosine distance
\begin{align}
d_{\cos}(\mathbf{a},\mathbf{b})
=1-\frac{\mathbf{a}^{\mathsf T}\mathbf{b}}
{\|\mathbf{a}\|_2\|\mathbf{b}\|_2}.
\label{eq_cos_dist}
\end{align}
This choice focuses the embedding on the shape of the activity pattern across neurons or electrodes, rather than on the absolute population firing magnitude alone. It is therefore closer to asking whether two time windows recruit the same latent population mode, while still allowing the overall size and contraction of each condition-specific cloud to be measured explicitly.

\section{Visual informatics methods}

\subsection{Principal Component Analysis (PCA)}
PCA is a widely used dimensionality reduction technique across various fields, including neuroscience. The core idea of PCA is to linearly transform the original high-dimensional data into a set of orthogonal components, known as principal components, in order to capture the maximum variance present in the data.

Mathematically, consider a transformation matrix $\boldsymbol{W}$ of size $N \times M$ that projects the data matrix $\boldsymbol{U}$ onto a lower-dimensional space $\mathbb{R}^M$, resulting in principal component scores $\boldsymbol{V}$. The transformation is expressed as:
\begin{align*}
    \boldsymbol{V} = \left( \boldsymbol{U} - \boldsymbol{U}' \right) \boldsymbol{W},
\end{align*}
where $\boldsymbol{U}'=\boldsymbol{1}_{R\mathcal{T}} \left\langle \boldsymbol{U} \right\rangle_{r,\tau}^T$ represents a $R\mathcal{T}\times N$ matrix, where each column $i$ stores the mean of the data across experimental conditions and time windows for each neuron $i$. Each principal component $\boldsymbol{V}_{(l)}$ for $l=1,\ldots, M$ is computed to successively inherit the maximum variance from the original data $\boldsymbol{U}$. The $k$-th transformation vector $\boldsymbol{W}_{(k)}$ is determined by maximizing the variance of the $k$-th principal component score while ensuring orthogonality to the previous $k-1$ transformation vectors:
\begin{align}
    \boldsymbol{W}_{(k)}&=\!\underset{\left\{ \!\substack{w:\|w\|=1\\
    w\perp w_{(l)},\,l=1,\ldots,k-1
    }
    \!\right\} }{\text{argmax}}\left\Vert \left(\boldsymbol{U}-\boldsymbol{U}'\right)\boldsymbol{w}\right\Vert ^{2}\nonumber\\&=\!\underset{\left\{ \!\substack{w:\|w\|=1\\
    w\perp w_{(l)},\,l=1,\ldots,k-1
    }
    \!\right\} }{\text{argmax}}\!\!\left\{ \!\boldsymbol{w}^{T}\!\!\left(\!\boldsymbol{U}\!-\boldsymbol{U}'\right)^{\!T}\!\left(\!\boldsymbol{U}\!-\!\boldsymbol{U}'\right)\boldsymbol{w}\!\right\} \nonumber\\&=\!\underset{\left\{ \boldsymbol{w}:\boldsymbol{w}\perp\boldsymbol{w}_{(l)},\ l=1,\ldots,k-1\right\} }{\text{argmax}}\left\{ \frac{\boldsymbol{w}^{T}\Sigma\boldsymbol{w}}{\boldsymbol{w}^{T}\boldsymbol{w}}\right\} ,
\end{align}
where $\Sigma = \left(\boldsymbol{U} - \boldsymbol{U}'\right)^{T} \left(\boldsymbol{U} - \boldsymbol{U}'\right)$ is the covariance matrix of the original data. Solving this optimization problem is equivalent to finding the eigenvectors of $\Sigma$, where $\boldsymbol{W}_{(k)}$ corresponds to the eigenvector associated with the $k$-th largest eigenvalue of $\Sigma$.

Despite its widespread use, PCA has several limitations when applied to neuronal activity data. One significant limitation is that principle components in PCA are assumed to be orthogonal. While extensions that try to relax this assumption to other forms of parametric dependence (i.e. linear in factor analysis techniques \cite{kim1978introduction}; specific non-linearities can be captured via the kernel PCA extension \cite{scholkopf1997kernel} or Gaussian process latent variable extensions \cite{lawrence2003gaussian}) have been well studied, some of them are not suitable for 2-D or 3-D visualization and most of them assume all principle components or factors are a linear (i.e. or nonlinear) combination of all of the input data \cite{farooq2024adaptive}. Inherently, this assumes that the neuronal activity patterns are uniform over time and leads to poor interpretability of those components in the presence of stimuli or other time-dependent neuronal re-organization. 

\subsection{t-distributed Stochastic Neighbor Embedding (t-SNE)}

t-SNE is a powerful technique for mapping high-dimensional data onto a lower-dimensional space (often two or three dimensions) in a nonlinear manner. Its primary objective is to place similar data points close to one another with high probability and to separate dissimilar points with high probability.  

Formally, let \(\boldsymbol{U}_i\) and \(\boldsymbol{U}_j\) be data points in the original high-dimensional space. The conditional probability  
\begin{align}
    p_{j|i} \;=\; \frac{\exp\!\bigl(-D(\boldsymbol{U}_i,\boldsymbol{U}_j)^2 / (2 \sigma_i^2)\bigr)}
    {\sum_{k \neq i} \exp\!\bigl(-D(\boldsymbol{U}_i,\boldsymbol{U}_k)^2 / (2 \sigma_i^2)\bigr)},
\end{align}  
captures how likely \(\boldsymbol{U}_i\) considers \(\boldsymbol{U}_j\) to be its neighbor, assuming neighbors follow a Gaussian distribution. The \emph{pairwise similarity} \(p_{ij}\) between \(\boldsymbol{U}_i\) and \(\boldsymbol{U}_j\) is then defined by averaging \(p_{j|i}\) and \(p_{i|j}\):
\begin{align}
    p_{ij} \;=\; \frac{p_{j|i} + p_{i|j}}{2 R \mathcal{T}}.
\end{align}

Let \(\boldsymbol{V}_i\) denote the projection of \(\boldsymbol{U}_i\) in the lower-dimensional space. t-SNE defines the similarities \(q_{ij}\) between \(\boldsymbol{V}_i\) and \(\boldsymbol{V}_j\) via a Student-\(t\) distribution:
\begin{align}
    q_{ij} \;=\; \frac{\bigl(1 + \|\boldsymbol{V}_i - \boldsymbol{V}_j\|^2\bigr)^{-1}}
    {\sum_{k \neq l} \bigl(1 + \|\boldsymbol{V}_k - \boldsymbol{V}_l\|^2\bigr)^{-1}}.
\end{align}
The embedding \(\{\boldsymbol{V}_i\}\) is then found by \emph{minimizing} the Kullback–Leibler (KL) divergence between \(P\) and \(Q\):
\begin{align}
    KL(P \,\|\, Q) \;=\; \sum_{i \neq j} p_{ij} \,\ln\!\Bigl(\tfrac{p_{ij}}{q_{ij}}\Bigr).
\end{align}

Most commonly, \(D(\boldsymbol{U}_i, \boldsymbol{U}_j)\) is taken to be the Euclidean distance. While natural in many settings, this choice can be problematic for neuronal data, which often live in very high-dimensional and \emph{sparsely occupied} spaces. In high-dimensional spaces, the ratio between the smallest pairwise distance and the largest pairwise distance often approaches 1, causing distances to \emph{concentrate}. In other words, points end up seeming roughly the same distance from one another, limiting the discriminative power of Euclidean distance. Moreover, because Euclidean distance treats all dimensions equally, large deviations in just a few coordinates can overshadow more subtle but meaningful patterns. As a result, Euclidean distance can fail to capture nonlinear or context-specific relationships in neuronal data, potentially obscuring important structures in the resulting visualization. Neuronal data often exhibits complex, non-linear relationships that Euclidean distance fails to account for, potentially leading to oversimplified interpretations of the underlying dynamics. In the results section, we will demonstrate the importance of selecting appropriate metrics to capture meaningful insights from the data.

Unlike PCA, which seeks to capture global variance directions, t-SNE emphasizes preserving \emph{local} structures by modeling similarities based on pairwise distances. This makes t-SNE particularly effective for revealing \emph{clusters} and patterns in neuronal activity—valuable for classifying responses under different experimental conditions. However, t-SNE’s focus on local neighborhoods can distort \emph{global} relationships, sometimes causing well-separated clusters to appear artificially close. Consequently, while t-SNE excels at highlighting \emph{localized} structure, it may miss overarching network dynamics or temporal continuity, limiting a comprehensive understanding of neuronal interactions over time.

\subsection{Minimum-Distortion Embedding (MDE)}

MDE is a general framework for dimensionality reduction \cite{agrawal2021minimum} that unifies popular techniques such as PCA, t-SNE, and UMAP \cite{pearson1901liii, van2008visualizing, mcinnes2018umap}. Consider an embedding function \(f\) that maps a high-dimensional data point \(\boldsymbol{U}_i\) to a lower-dimensional point \(\boldsymbol{V}_i\), i.e.\ \(f(\boldsymbol{U}_i) = \boldsymbol{V}_i\). Let \(\delta_{ij}\) be the target distance between \(\boldsymbol{U}_i\) and \(\boldsymbol{U}_j\) in the original space. For MDE with the cosine metric, we set \(\delta_{ij}=d_{\cos}(\boldsymbol{U}_i,\boldsymbol{U}_j)\). For the Euclidean comparison, we instead set \(\delta_{ij}=\|\boldsymbol{U}_i-\boldsymbol{U}_j\|_2\). In both cases, \(d_{ij}=\|\boldsymbol{V}_i-\boldsymbol{V}_j\|_2\) denotes the distance in the two-dimensional embedding. Rather than minimizing a simple ratio (which trivially leads to collapsed points), MDE aims to preserve original distances by minimizing an average distortion loss that penalizes both expansion and contraction:
\begin{align}\label{eq:mde-distortion}
	\mathcal{E} \;=\;
	\frac{1}{C_{2}^{R\mathcal{T}}} \sum_{i < j}
	\ell\bigl(d_{ij}, \delta_{ij}\bigr),
\end{align}

where \(C_{2}^{R\mathcal{T}}\) represents the total number of unique pairs in the dataset, and \(\ell(\cdot, \cdot)\) is a specified penalty function, here we choose \(\ell(\cdot, \cdot)\) as the fractional relative loss \(\ell(d_{ij}, \delta_{ij}) = (d_{ij} - \delta_{ij})^2 / \delta_{ij}^2\). The goal of MDE is to \emph{minimize} \(\mathcal{E}\) via gradient-based optimization with respect to the embedded coordinates \(\{\boldsymbol{V}_i\}\).

By selecting specific penalty functions and weightings in \eqref{eq:mde-distortion}, MDE recovers various traditional dimensionality reduction methods. For instance, PCA \cite{pearson1901liii} can be framed as a linear special case of MDE under Euclidean geometry by selecting a loss function derived from inner products. Conversely, t-SNE \cite{van2008visualizing} can be viewed as a locally focused variant of MDE, where the distortion function heavily penalizes the disruption of nearest-neighbor relationships while global distances are deemphasized.

In this work, we do \emph{not} assume a specific parametric form for \(f\) and directly optimize \eqref{eq:mde-distortion} to find the embedded coordinates. Unless explicitly stated as the Euclidean comparison, MDE and t-SNE use \(d_{\cos}\) as the distance between population activity vectors. In neuronal firing-rate data, absolute magnitudes can be less informative than the direction, or shape, of the population activity vector, so using cosine distance better highlights functional relationships among neurons. This choice often preserves global structure more effectively because even data points with similar population-activity directions still provide meaningful relational information, avoiding the ``crowding'' pitfalls that can arise from purely Euclidean assumptions in high dimensions.

Thus, although our MDE implementation remains sensitive to local structure, the cosine-based metric naturally incorporates global patterns by leveraging the fact that most points share some level of directional similarity in population space. In practice, this metric choice offers a promising trade-off between local fidelity and global integrity, mitigating the distortions and crowding effects more commonly seen when relying on Euclidean distances for embedding high-dimensional neuronal data.

\section{Results}
We employ visual informatics techniques to study two separate scenarios, namely the development of cortical neuronal cultures over time, and the changes in neuronal activities in response to stimulation. In particular, we conducted both \textit{in silico} and \textit{in vitro} experiments for the two scenarios. We compare the results obtained using PCA, t-SNE, and MDE, employing different selections of metrics. Our findings demonstrate that we can obtain the trajectory of neuronal culture development over time using MDE, and that we can identify the different phases before and after the application of the stimulation protocol.

\subsection{Development of \textit{in silico} neuronal network}\label{sec:inscilicodevelopment}
\begin{figure*}
	\centering
	\begin{tabular}{@{}c@{\hspace{0.006\linewidth}}c@{\hspace{0.006\linewidth}}c@{}}
        \begin{tabular}[t]{@{}l@{}}A\\[-0.25em]
        \raisebox{-10em}[0pt][0pt]{\includegraphics[width=0.25\linewidth, trim=0 0 0 0, clip]{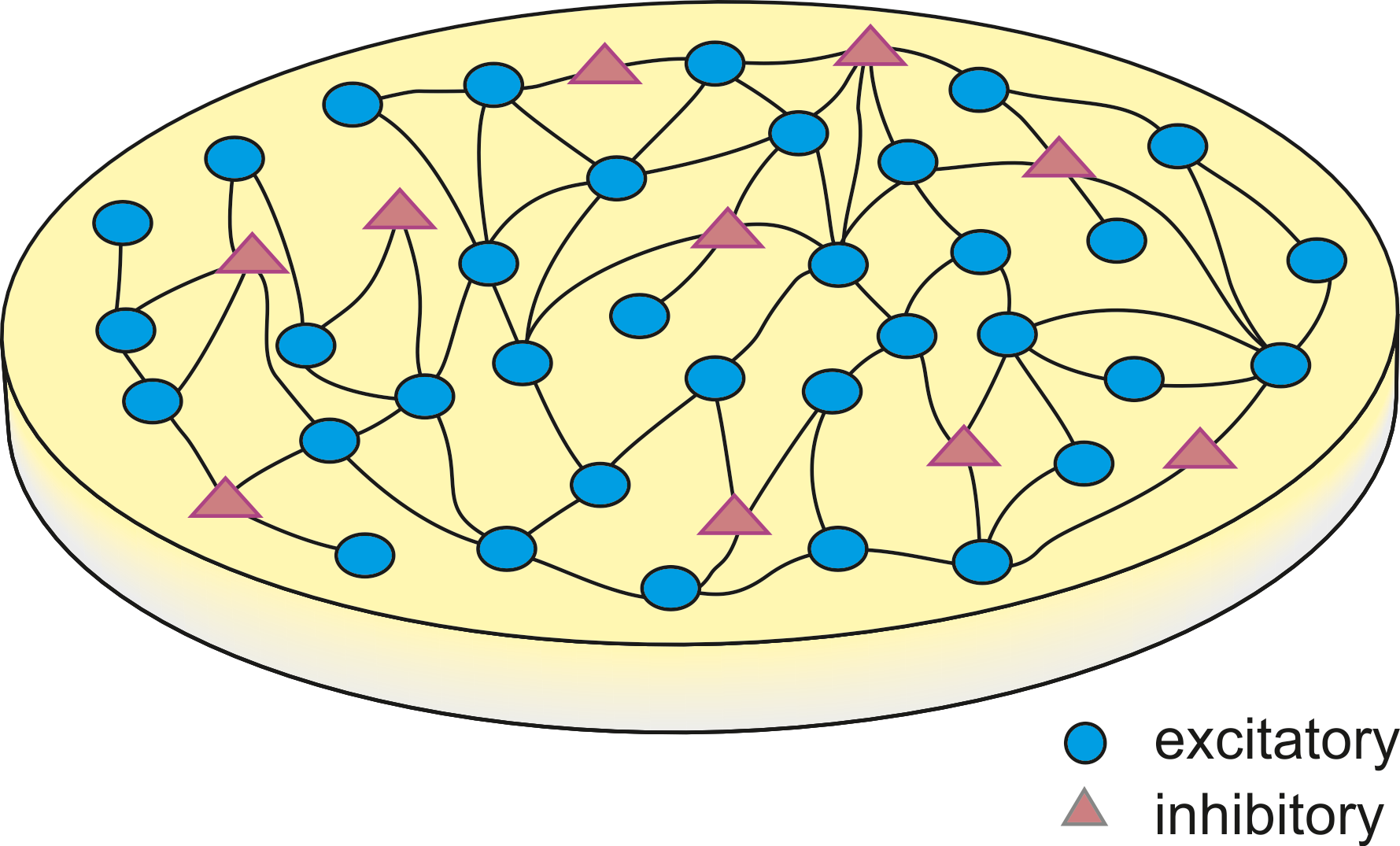}}\end{tabular}
        &
        \begin{tabular}[t]{@{}l@{}}B\\[-0.25em]
        {\includegraphics[width=0.326\linewidth, trim=0 0 0 0, clip]{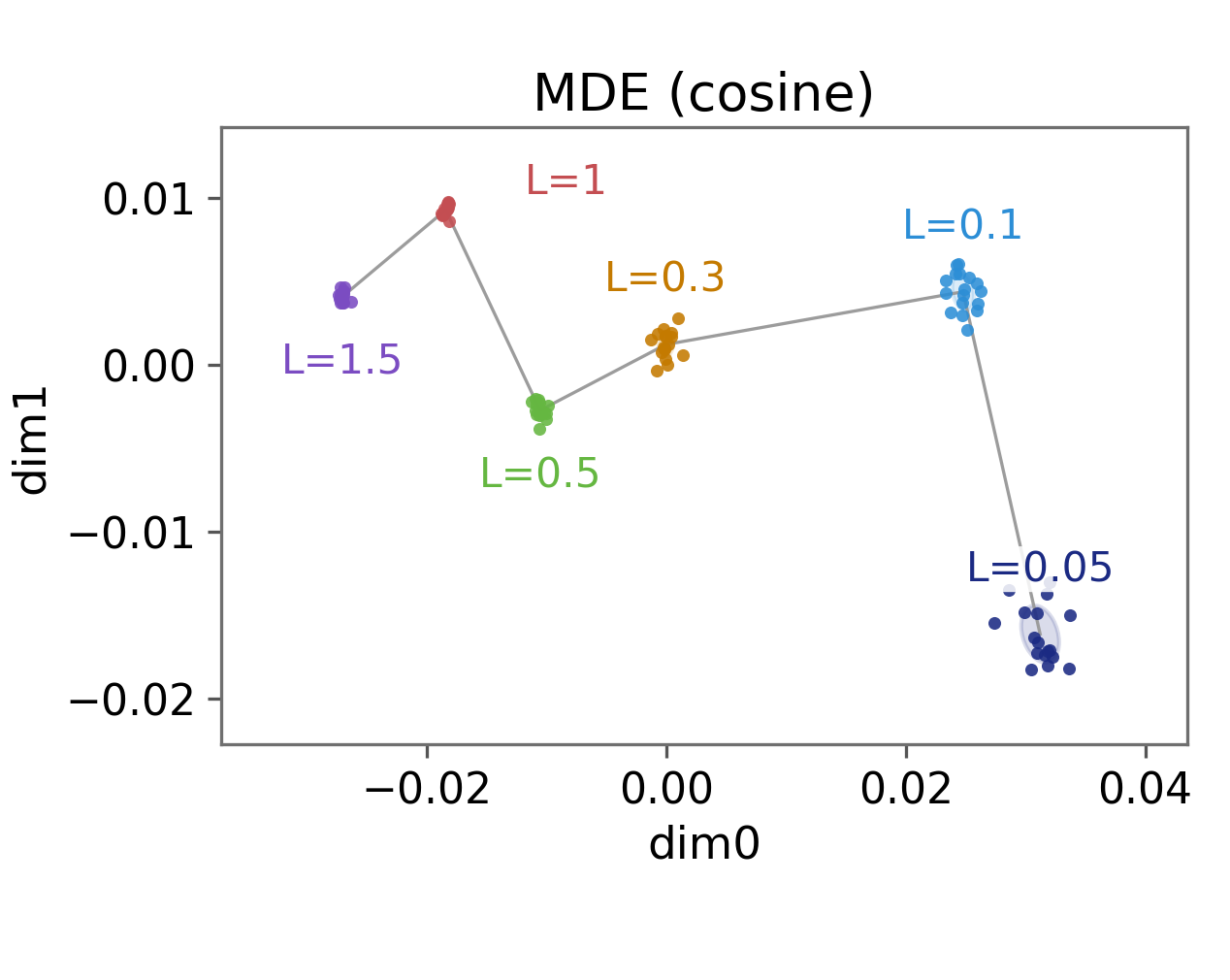}}\end{tabular}
        &
        \begin{tabular}[t]{@{}l@{}}C\\[-0.25em]
        {\includegraphics[width=0.326\linewidth, trim=0 0 0 0, clip]{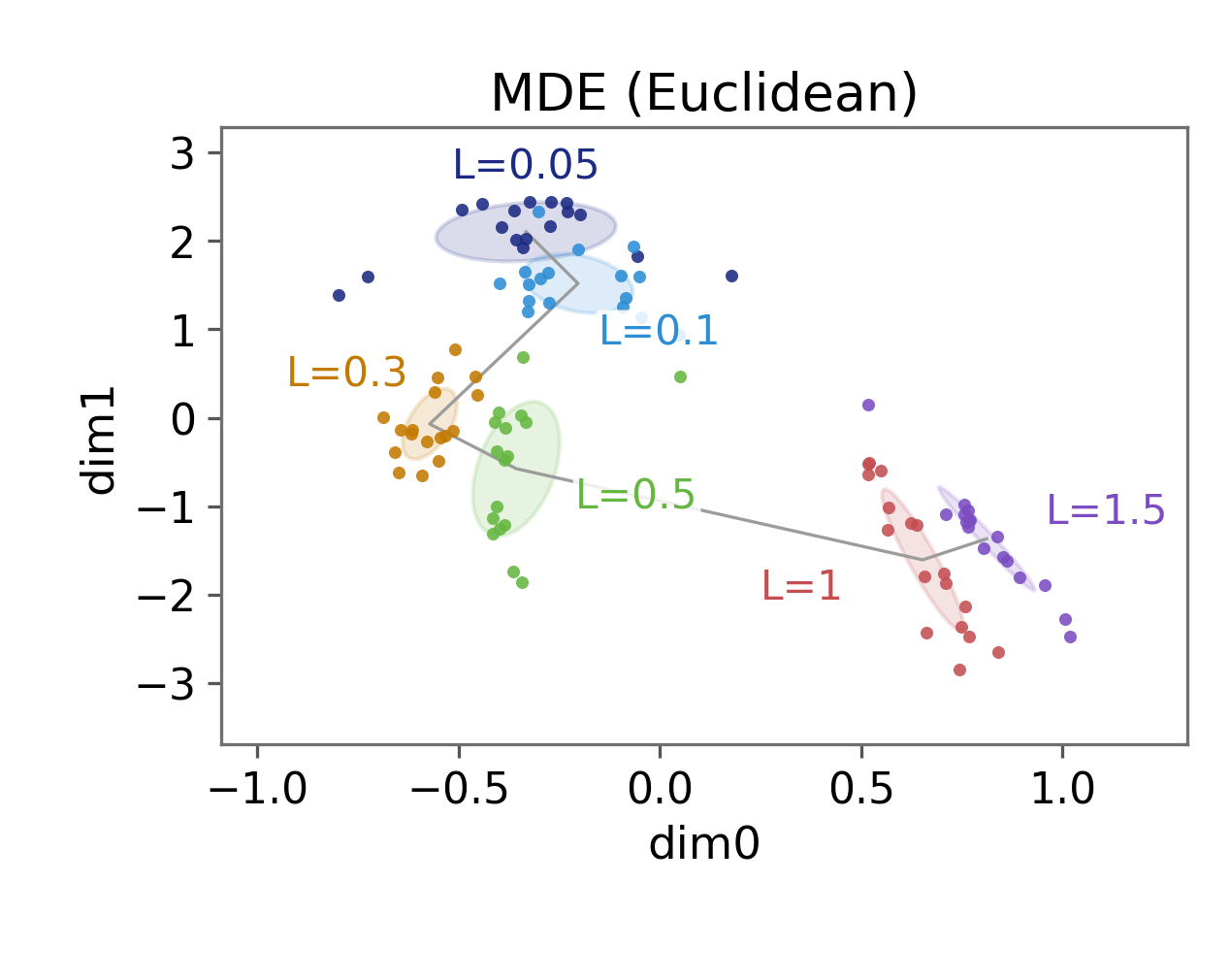}}\end{tabular} \\[-2.3em]
        \begin{tabular}[t]{@{}l@{}}D\\[-0.25em]
        {\includegraphics[width=0.326\linewidth, trim=0 0 0 0, clip]{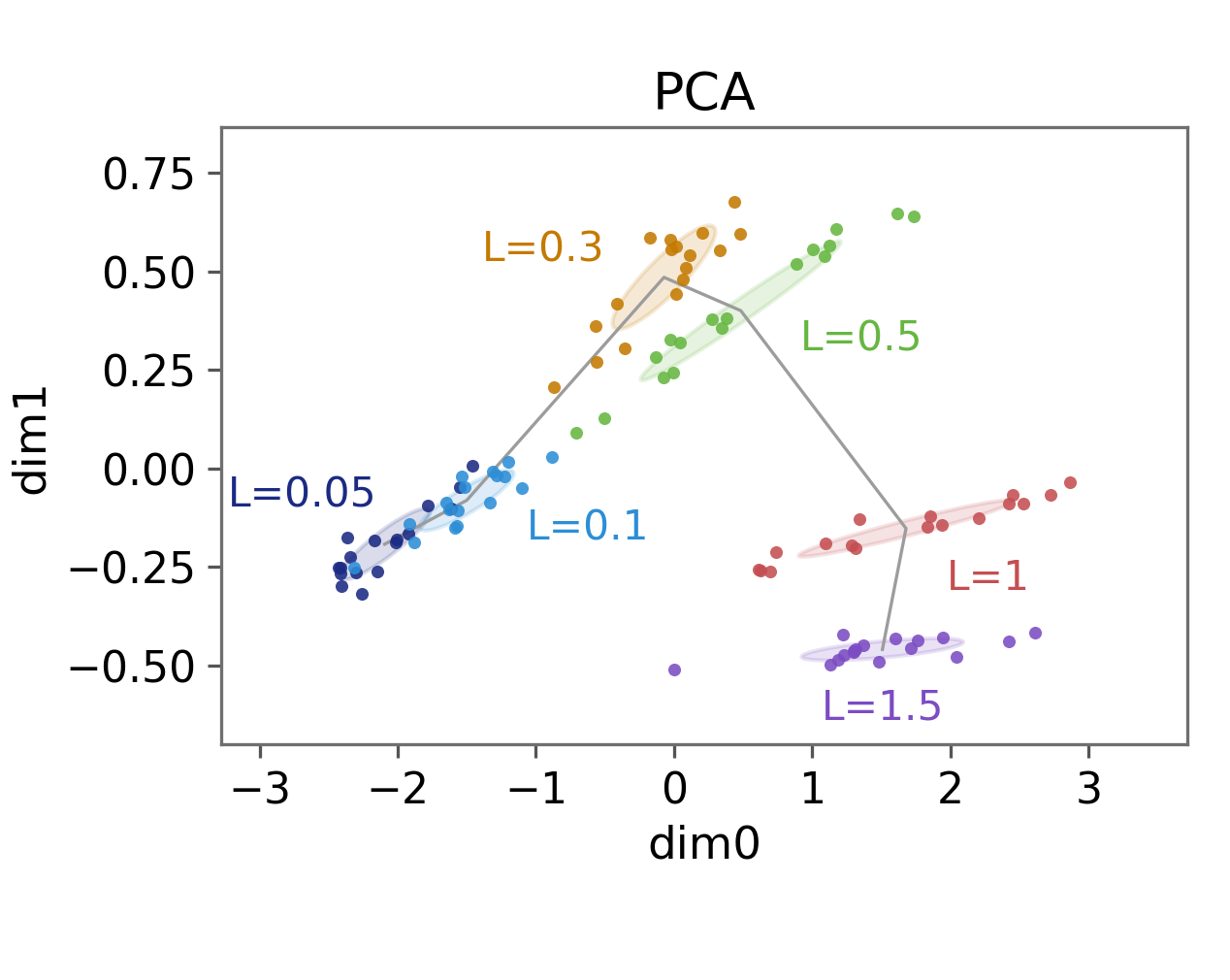}}\end{tabular}
        &
        \begin{tabular}[t]{@{}l@{}}E\\[-0.25em]
        {\includegraphics[width=0.326\linewidth, trim=0 0 0 0, clip]{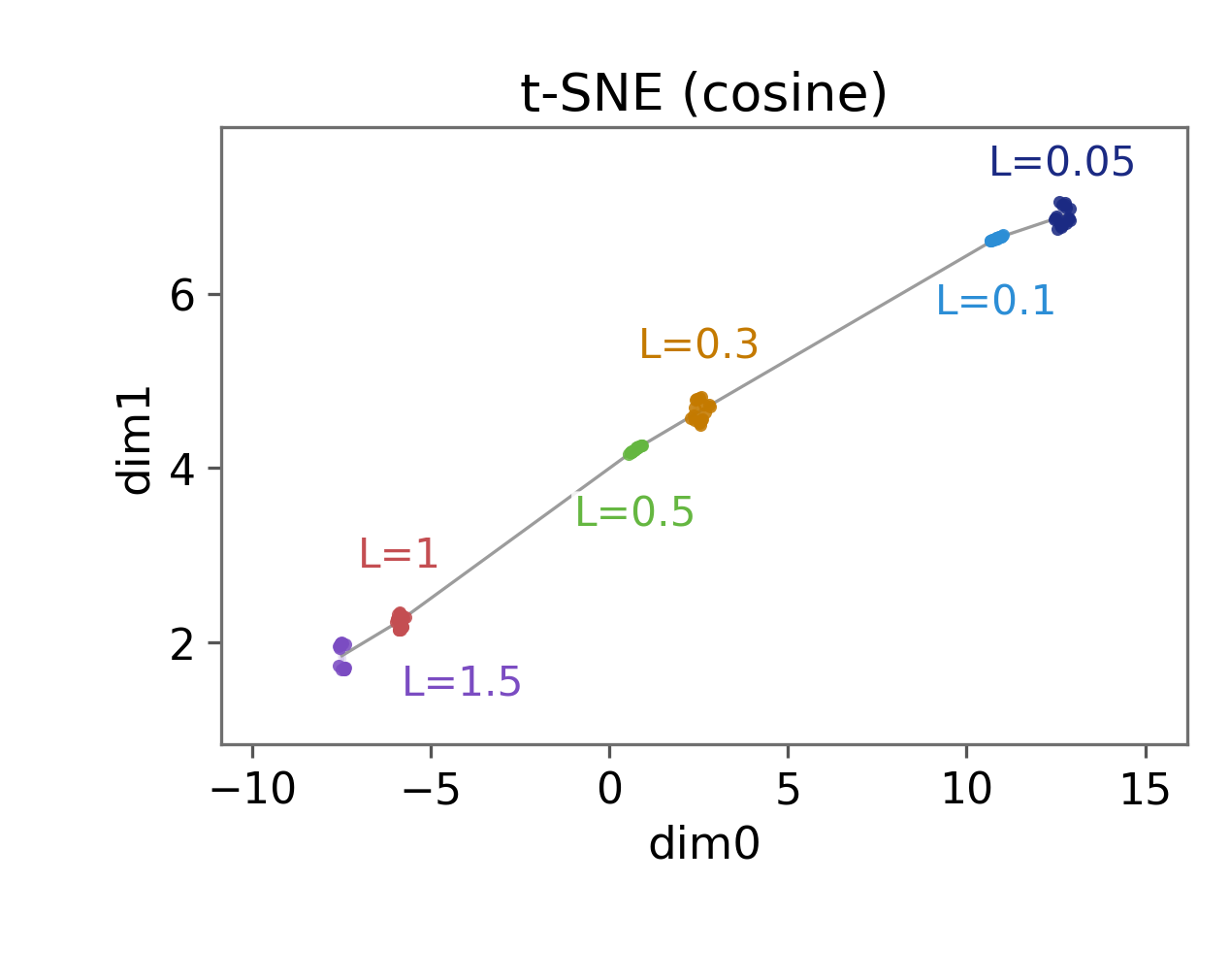}}\end{tabular}
        &
        \begin{tabular}[t]{@{}l@{}}F\\[-0.25em]
        {\includegraphics[width=0.326\linewidth, trim=0 0 0 0, clip]{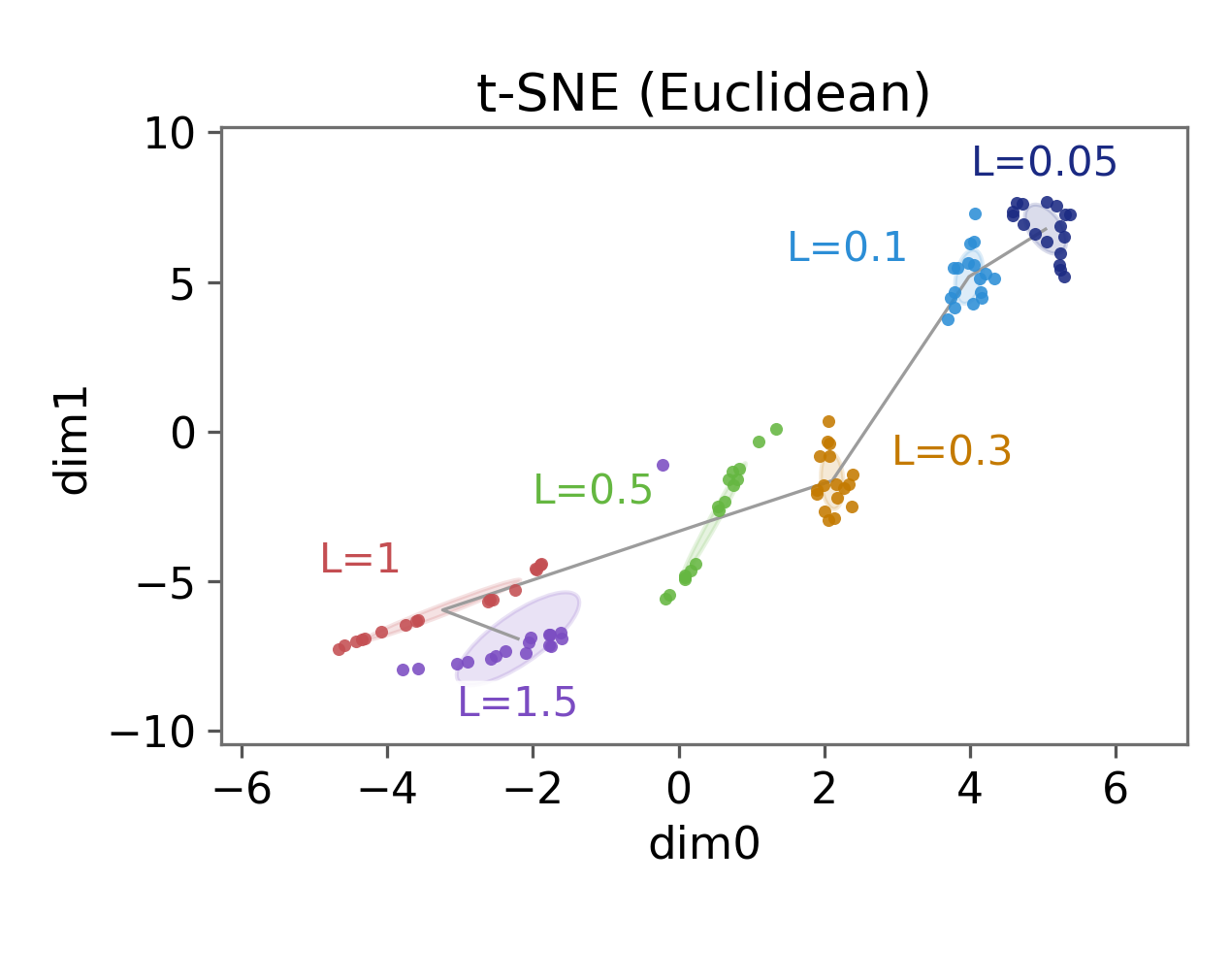}}\end{tabular}\\[-2.3em]

        \begin{tabular}[t]{@{}l@{}}G\\[-0.25em]
        {\includegraphics[width=0.326\linewidth, trim=0 0 0 0, clip]{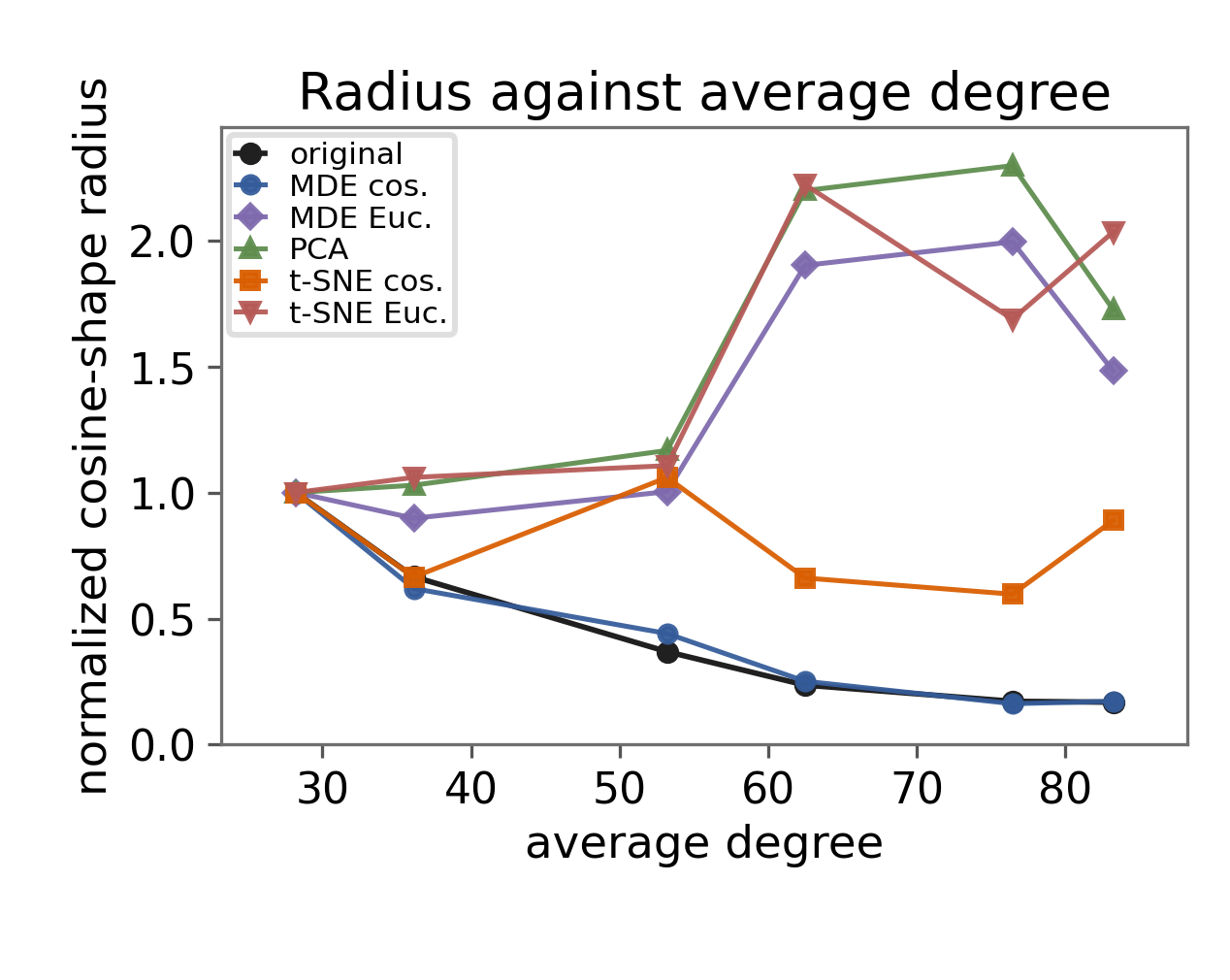}}\end{tabular}
        &
        \begin{tabular}[t]{@{}l@{}}H\\[-0.25em]
        {\includegraphics[width=0.326\linewidth, trim=0 0 0 0, clip]{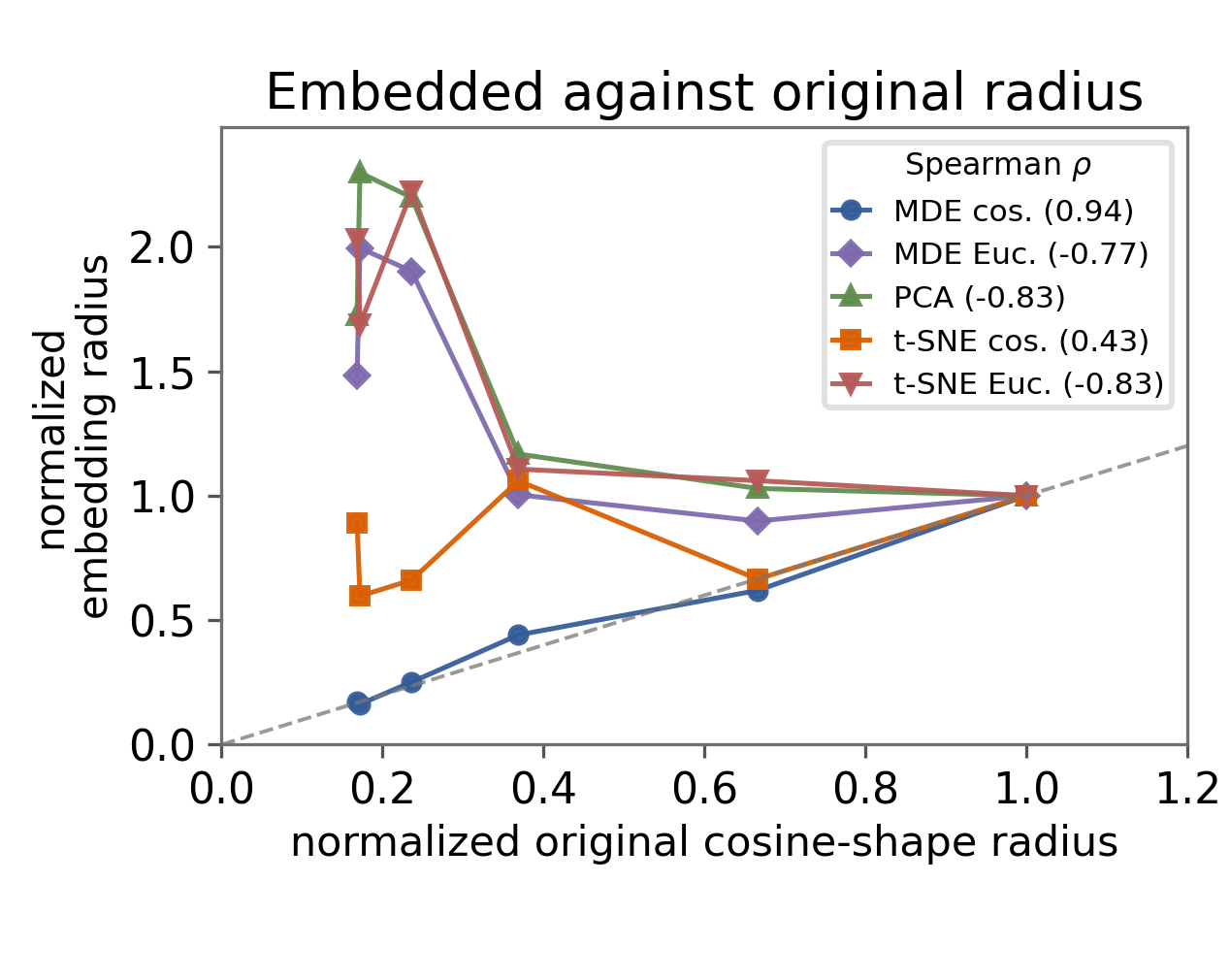}}\end{tabular}
        &
        \begin{tabular}[t]{@{}l@{}}I\\[-0.25em]
        {\includegraphics[width=0.326\linewidth, trim=0 0 0 0, clip]{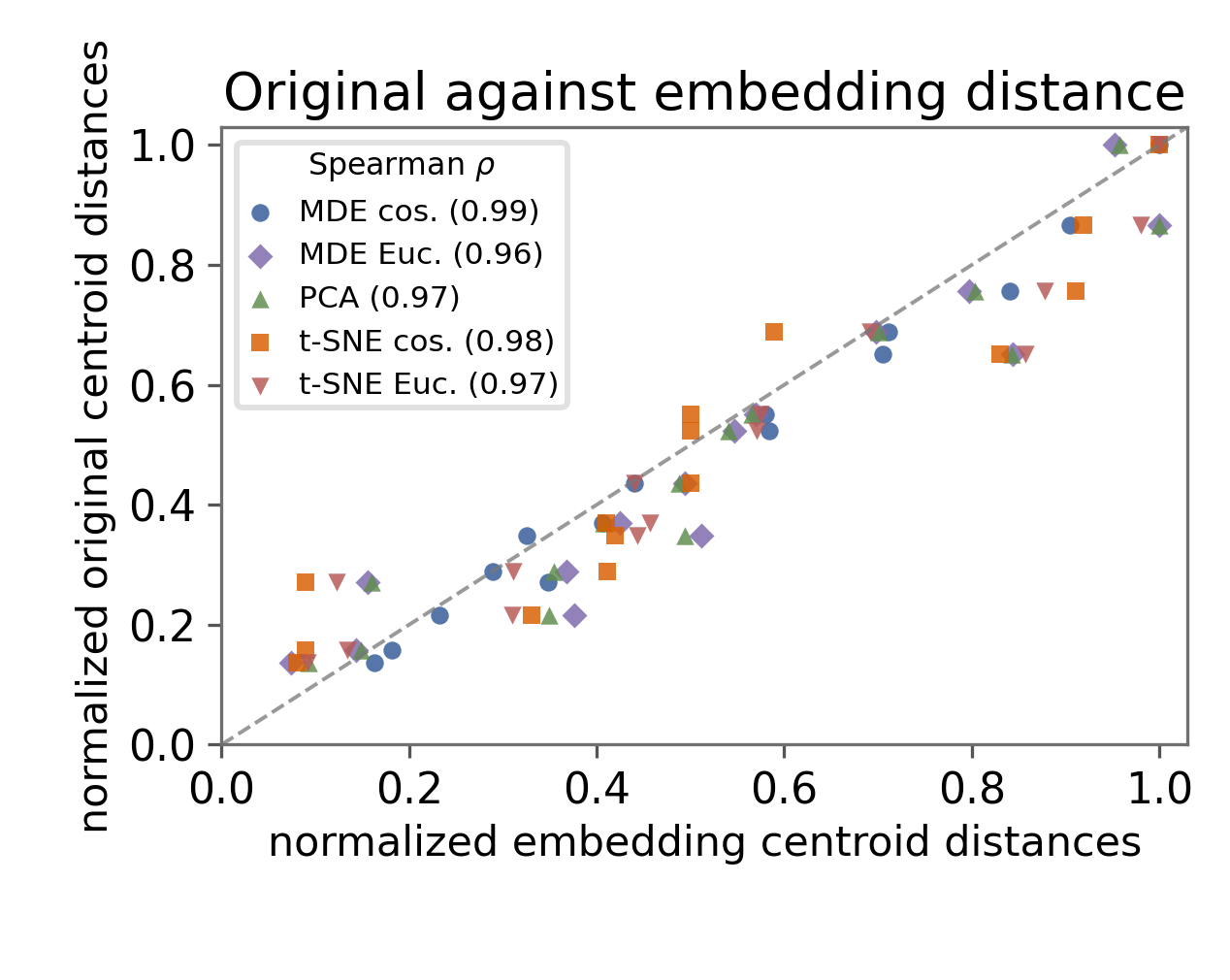}}\end{tabular}
	\end{tabular}
	\caption{
	   {A:} A three-dimensional illustration of the \emph{in silico} experimental setup. In this configuration, the neurons (totaling $N=195$) are initially arranged on a flat surface in a homogeneous manner. Neuronal activities are simulated using snapshot network topologies characterized by specific axon lengths \(L\). {B-F:} Visualization of neuronal activity embeddings derived from different snapshot topologies with varying axon lengths. The embeddings are generated using B: MDE with a cosine metric, C: MDE with a Euclidean metric, D: PCA, E: t-SNE with a cosine metric, and F: t-SNE with a Euclidean metric, respectively. 
       In all plots, each dot corresponds to the average firing rates of neurons $U^{r}(\tau)$ in a time window of $50s$ in a network simulation for a given a state of the network, with color shadings providing a guide to the eye for the clustering of the data.
       Dots of the same color correspond to the same simulation run. G shows the normalized comparison between the original-space cosine-shape radius and the displayed embedding cluster size as a function of the average network degree. H shows the normalized embedding radius against the normalized original cosine-shape radius, where the dashed line denotes equality between original and embedded cloud contraction. I shows the normalized original centroid distances against the normalized embedding centroid distances for all pairs of maturation conditions.}
	\label{fig_silico_dev}
\end{figure*}

Given the challenges and costs associated with validating neuronal growth and development through \emph{in vitro} experiments, we first employ an \emph{in silico} model, as shown in \fig{fig_silico_dev}A, to simulate the growth dynamics of a neuronal system~\cite{houben2025role} consisting of \(N=195\) neurons with zero initial connectivity. We simulate culture development by increasing the axon length \(L\) from \(0.05\) to \(1.5\)~mm, so that connections are gradually established and the average network degree increases accordingly. We take snapshots of the network topology at different values of \(L\), use each topology to simulate spontaneous neuronal activities, and analyze the resulting activity data using MDE, PCA, and t-SNE with different choices of distance metrics.

To compare these methods, we quantify how well each embedding preserves the size and relative position of the activity patterns. For each value of \(L\), let \(\mathbf{u}_{L,i}\) be the firing-rate vector of time window \(i\), and let \(\bar{\mathbf{u}}_{L}=n_L^{-1}\sum_i\mathbf{u}_{L,i}\) be the condition-level activity pattern. We define the original cosine-shape radius as
\begin{align}
    R_{\mathrm{orig}}(L)
    &=\frac{1}{n_L}\sum_{i=1}^{n_L}
    d_{\cos}\!\left(\mathbf{u}_{L,i},\bar{\mathbf{u}}_{L}\right),
\label{eq:orig_radius}
\end{align}
Here \(d_{\cos}\) is the cosine distance defined in \req{eq_cos_dist}. Thus, \(R_{\mathrm{orig}}(L)\) measures dispersion in population-pattern direction rather than firing-rate magnitude. For an embedding method \(m\), with embedded points \(\mathbf{z}^{(m)}_{L,i}\), we compute the corresponding embedding radius by
\begin{align}
    R^{(m)}_{\mathrm{emb}}(L)
    &=\frac{1}{n_L}\sum_{i=1}^{n_L}
    \left\|\mathbf{z}^{(m)}_{L,i}-\bar{\mathbf{z}}^{(m)}_{L}\right\|_2,
\label{eq:emb_radius}
\end{align}
where $\bar{\mathbf{z}}^{(m)}_{L}=n_L^{-1}\sum_i\mathbf{z}^{(m)}_{L,i}$ is the embedded centroid of condition \(L\). For the relative positions, we compare the pairwise condition-centroid distances in the original space with those in the embedding,
\begin{align}
    D_{\mathrm{orig}}(L,L')
    &=d_{\cos}\!\left(\bar{\mathbf{u}}_{L},\bar{\mathbf{u}}_{L'}\right),
\label{eq:centroid_distance}
\end{align}
where $D^{(m)}_{\mathrm{emb}}(L,L')=\left\|\bar{\mathbf{z}}^{(m)}_{L}-\bar{\mathbf{z}}^{(m)}_{L'}\right\|_2$ is the embedded centroid distance. These quantities test whether the visualization preserves both activity-cloud contraction and distances between maturation states.

\subsubsection{Performance of PCA}
\fig{fig_silico_dev}D illustrates the results obtained using PCA. The data points generated from the same topology are more scattered than those obtained using MDE and t-SNE, and the clusters corresponding to different \(L\) values are less clearly separated. This is expected because PCA only searches for orthogonal directions of maximal variance, and therefore provides a linear summary of the firing-rate data. Such a projection can capture a dominant monotone component of the simulated maturation process, but it does not directly preserve the nonlinear similarity structure between activity patterns.

This limitation can be seen more clearly from \fig{fig_silico_dev}G and \fig{fig_silico_dev}H. The original cosine-shape radius decreases with the average network degree, indicating that the simulated activity cloud contracts as the network matures. However, PCA does not reproduce this contraction in the embedding. It can make mature conditions appear visually larger even though the original activity-pattern cloud becomes smaller. On the other hand, \fig{fig_silico_dev}I shows that PCA performs well for the centroid-distance diagnostic. This is reasonable because the simulated maturation coordinate is approximately one-dimensional and gives a strong monotone variance direction. Nevertheless, preserving the centroid distances alone is not sufficient, because the embedding should also preserve how the activity cloud changes with maturation. This limitation is important for the \emph{in vitro} analysis, where PCA does not reveal the structured developmental trajectory shown in \fig{fig_vitro_dev}.

\subsubsection{Effectiveness of MDE and t-SNE}
In contrast, both MDE and t-SNE classify the neuronal activities more clearly, with distinct clusters emerging for different snapshot topologies. As shown in \fig{fig_silico_dev}B, MDE with the cosine metric captures a clear developmental trajectory. The clusters are ordered as \(L\) increases, and this ordering reflects the gradual increase of connectivity within the simulated network. More importantly, \fig{fig_silico_dev}G and \fig{fig_silico_dev}H show that MDE with the cosine metric is the only method whose embedded radius closely follows the contraction of the original cosine-shape radius. This indicates that the embedding preserves not only the relative position of the developmental states, but also the reduction in activity-pattern variability as the network matures.

t-SNE with the cosine metric also separates the activities into distinct clusters, as shown in \fig{fig_silico_dev}E. This confirms that t-SNE is effective at identifying local group structure in the data. However, the positions of the t-SNE clusters are less informative as a developmental coordinate. The clusters are well separated, but their relative distances and sizes do not follow the original cosine-shape radius as well as MDE. Therefore, t-SNE is useful for classification of activity states, while MDE provides a more informative representation of both the local clusters and the global developmental geometry.

\subsubsection{Importance of Metric Selection}
The choice of distance metric is therefore crucial for both MDE and t-SNE. As demonstrated in \fig{fig_silico_dev}C and \fig{fig_silico_dev}F, the embeddings obtained using Euclidean distance are less effective. Although some local grouping can be observed, the developmental ordering and the relation between cluster size and original activity-pattern variability are distorted. This is because Euclidean distance is sensitive to absolute firing-rate magnitude, while the developmental changes in this analysis are more naturally described by the shape of the population activity vector.

Cosine distance compares activity vectors by their direction in population space, and therefore focuses on whether two time windows group similar population activity patterns. This metric choice explains why both MDE and t-SNE improve when cosine distance is used. The advantage is strongest for MDE because MDE uses the pairwise geometry more directly to preserve distances globally. In contrast, t-SNE remains more locally focused, and PCA remains limited by its linear variance-based representation. Thus, the \emph{in silico} results suggest that the most informative visualization is obtained when the embedding method and the distance metric are both matched to the structure of neuronal population activity.

Together, these results show that the simulated maturation data are not only separated into different activity states, but also contain a measurable contraction of the population activity cloud. PCA captures part of the dominant maturation direction, and t-SNE captures local clusters, but MDE with cosine distance best preserves both the ordering of the developmental states and the change in activity-pattern variability. We therefore use this result as the basis for analyzing the biological developmental data.

\subsection{Development of \textit{in vitro} human cortical neuronal network}
Following our examination of various dimensionality reduction techniques and their effectiveness, we now apply these methods to an \emph{in vitro} human iPSC-derived cortical neuronal network (purely excitatory). The data encompasses neuronal activities recorded throughout the network's development from day \emph{in vitro} (DIV) 23 to DIV~64. Details of data
acquisition and analysis are provided in the Data Generation section. This experimental setup provides an exciting opportunity to investigate the dynamics of neuronal network development, allowing us to explore how connectivity and activity patterns change over time. By analyzing these complex interactions, we aim to gain a deeper understanding of the fundamental processes that govern neuronal behavior and the factors that influence network formation.

\begin{figure*}
    \centering
    \begin{tabular}{@{}c@{\hspace{0.006\linewidth}}c@{}}
        \begin{tabular}[t]{@{}l@{}}A\\[-0.25em]
        {\includegraphics[width=0.475\linewidth, trim=0 0 0 0, clip]{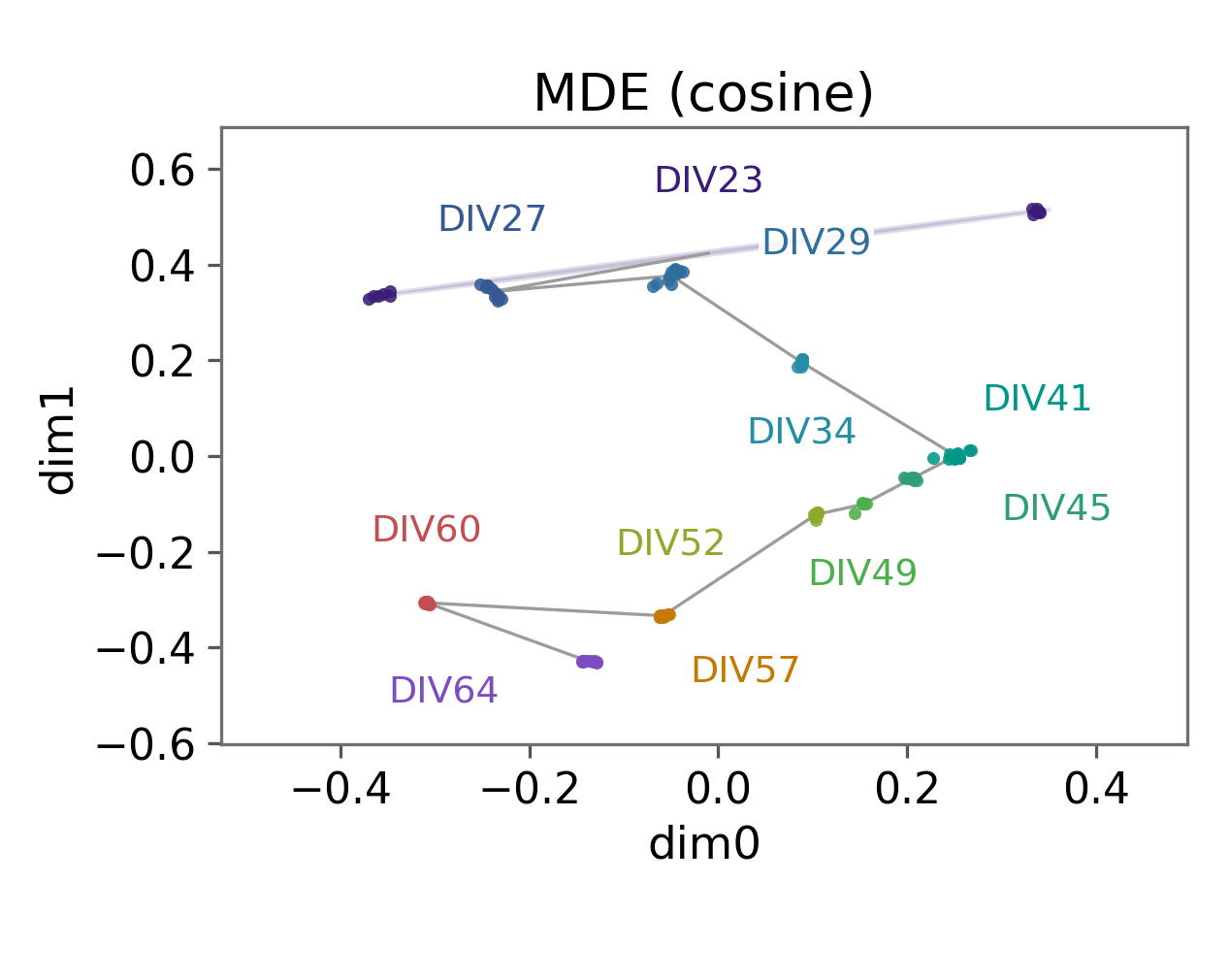}}\end{tabular}
        &
        \begin{tabular}[t]{@{}l@{}}B\\[-0.25em]
        {\includegraphics[width=0.475\linewidth, trim=0 0 0 0, clip]{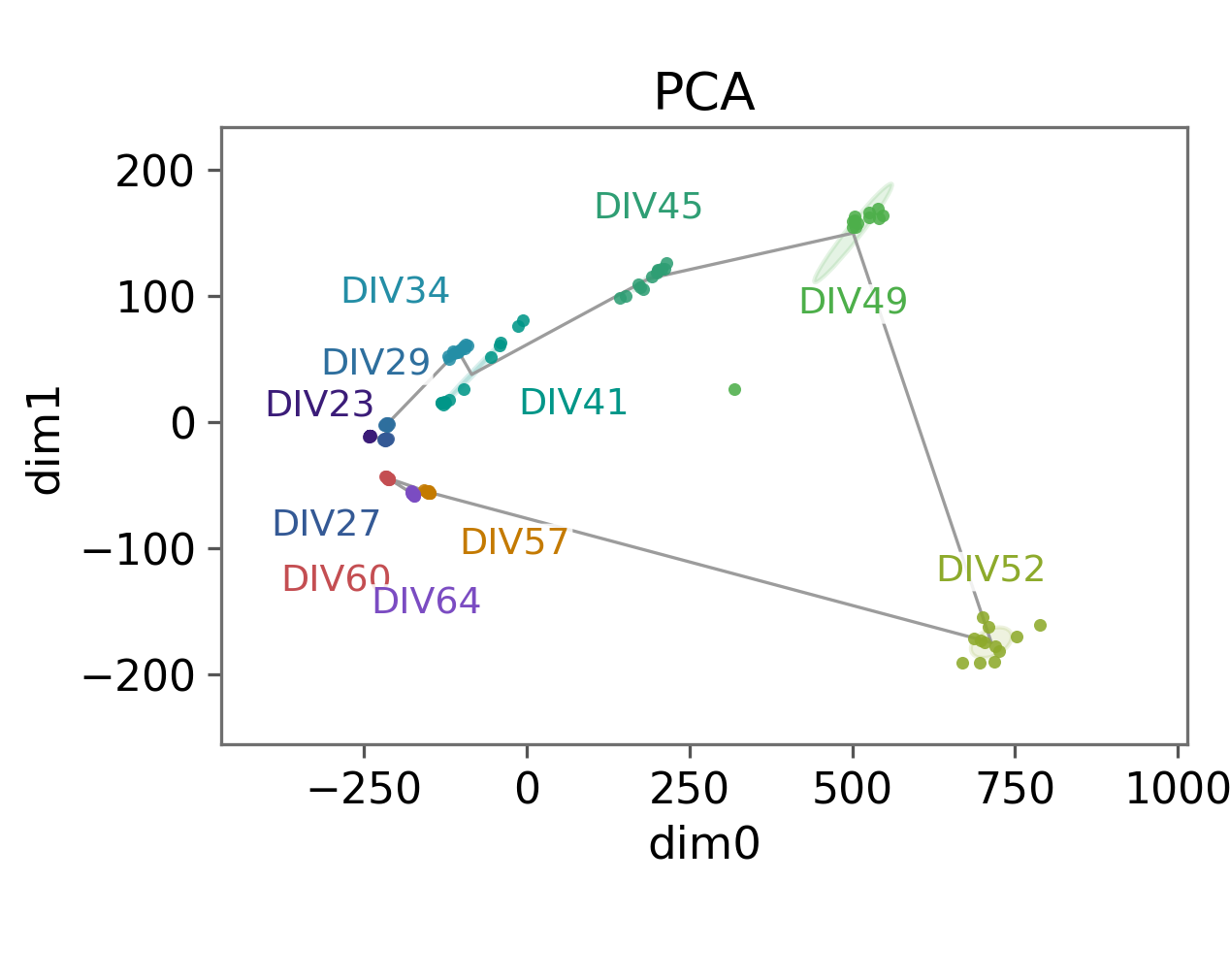}}\end{tabular}\\[-2.3em]
        \begin{tabular}[t]{@{}l@{}}C\\[-0.25em]
        {\includegraphics[width=0.475\linewidth, trim=0 0 0 0, clip]{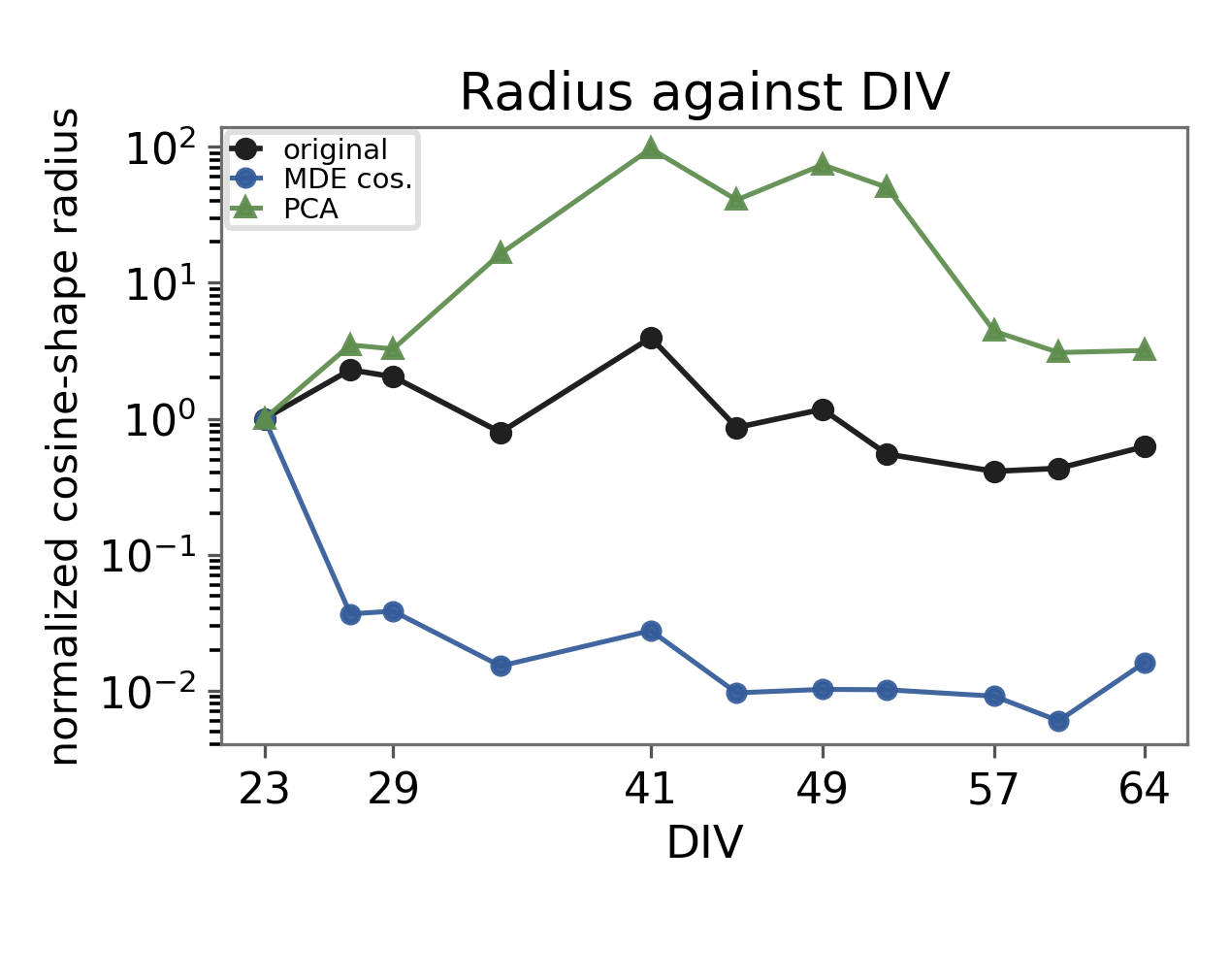}}\end{tabular}
        &
        \begin{tabular}[t]{@{}l@{}}D\\[-0.25em]
        {\includegraphics[width=0.475\linewidth, trim=0 0 0 0, clip]{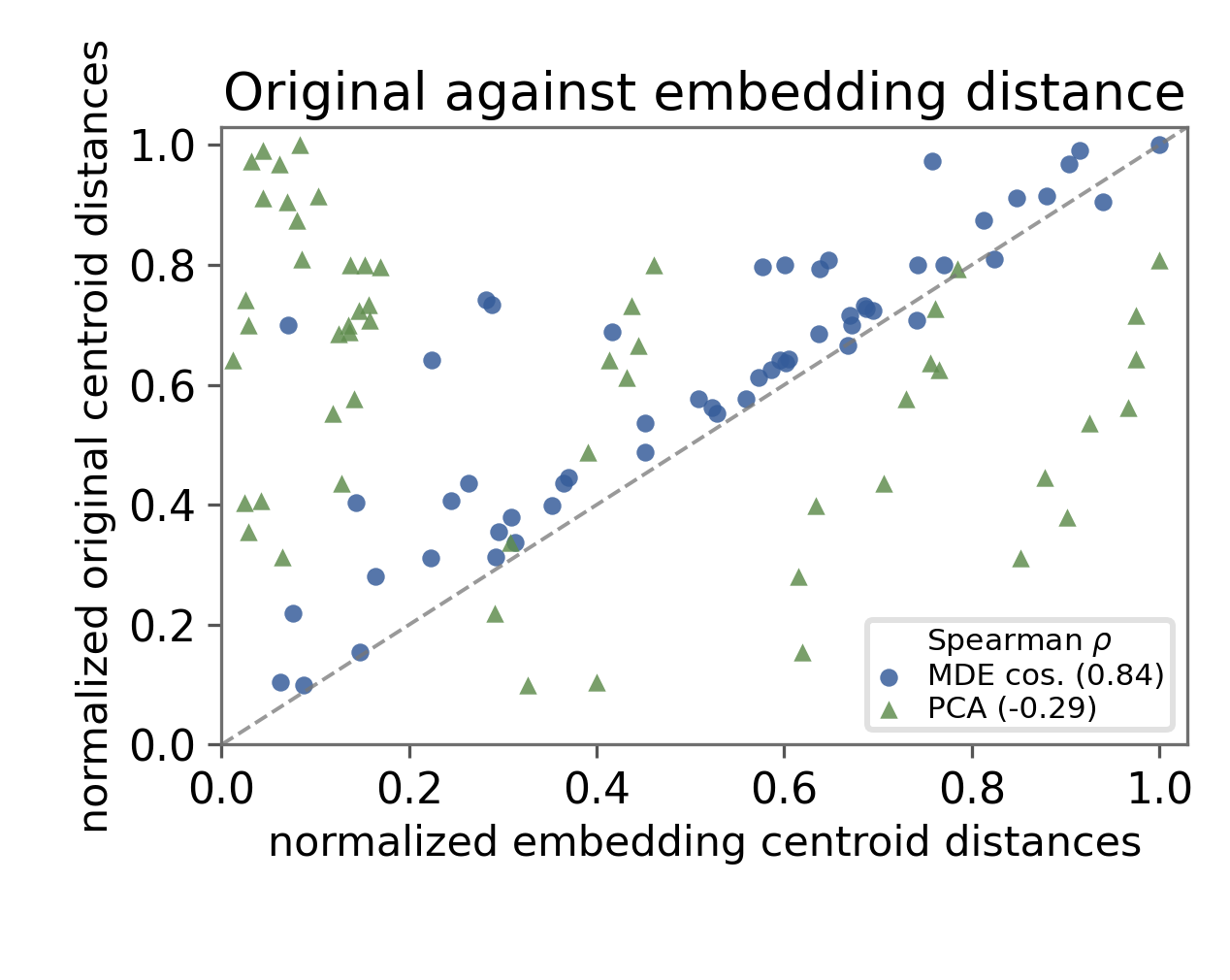}}\end{tabular}
    \end{tabular}
    \caption{Visualization and quantitative analysis of \emph{in vitro} human cortical neuronal activity recorded from DIV23 to DIV64. A and B: The embeddings are created using MDE with a cosine metric and PCA, respectively. Each dot corresponds to a $50s$ window of the recording, and colors indicate the DIV. C: Normalized original cosine-shape radius and normalized embedding radius against DIV for the original activity space, MDE with a cosine metric, and PCA. A logarithmic y-axis is used because the PCA radius is much larger than the original and MDE radii at intermediate DIVs. D: Normalized original centroid distances against normalized embedding centroid distances for all pairs of DIVs, comparing MDE with a cosine metric and PCA.}
    \label{fig_vitro_dev}
\end{figure*}

As shown in the \emph{in silico} study in section \ref{sec:inscilicodevelopment}, we observe that MDE with a cosine metric is better suited for preserving the global developmental geometry of neuronal activity. We therefore apply MDE with a cosine metric to the human iPSC-derived cortical activity, and compare the result with PCA. We also use the same quantitative diagnostics introduced above, namely the original cosine-shape radius, the embedding radius, and the pairwise DIV-centroid distances. This allows us to ask whether the two-dimensional map preserves not only the visual separation between DIVs, but also the contraction and relative displacement of the population activity patterns.

As shown in \fig{fig_vitro_dev}A, MDE separates the neuronal activity into a structured developmental trajectory. The activity at DIV23 is split into two local clusters, suggesting that the early culture can occupy two population activity modes within the same recording day, which may potentially be due to the functional heterogeneity of the early network state when the culture is immature. From DIV23 to DIV34, the centroids occupy an early branch of the embedding. Around DIV41, the trajectory changes direction and then follows a consistent direction towards the later DIVs. This is a plausible developmental geometry, because the late DIVs are separated from the early DIV23 state rather than returning to it. In contrast, PCA in \fig{fig_vitro_dev}B gives a less stable trajectory. The direction changes around DIV49 and DIV52, and the later DIVs from DIV57 are close to DIV23 in the PCA map. If interpreted directly, this would suggest that the culture moves back to an early activity state, which is not consistent with progressive maturation over the recording period. Therefore, the PCA embedding separates some individual DIVs, but the ordering of the DIV centroids is not a reliable representation of network development.

The cosine-shape radius in \fig{fig_vitro_dev}C provides a second view of the same transition. In the original activity space, the within-DIV radius first increases from DIV23 to DIV27-DIV29, decreases at DIV34, and reaches its largest value at DIV41. The radius then decreases at later DIVs, indicating that the population activity patterns become less dispersed after this transition. This suggests that DIV41 may correspond to a transient reorganization of the network activity, where the culture explores a broader set of population patterns before settling into a more stable collective regime. This interpretation is compatible with long-term studies showing that human stem-cell-derived neuronal networks can change their morphology, activity maps, bursting dynamics, and functional properties over weeks of development~\cite{frega2017rapid,habibey2022longterm}. MDE with a cosine metric preserves the small radii of the later DIVs and retains a local increase around DIV41. PCA, however, gives a very large embedding radius for the intermediate DIVs, especially around DIV41-DIV52. This large PCA radius does not agree with the original cosine-shape radius, and indicates that the spread in the PCA map is dominated by projection scale rather than the original activity-pattern dispersion.

The third diagnostic,is the pairwise DIV-centroid distance shown in \fig{fig_vitro_dev}D. MDE with a cosine metric preserves the ordering of pairwise DIV-centroid distances well, with a Spearman correlation of \(0.84\). PCA gives a negative correlation of \(-0.29\), showing that distances between DIVs in the PCA map do not reflect the distances between their activity centroids in the original space. Therefore, the main advantage of MDE is not only that it gives a visually clearer embedding, but that it preserves the developmental geometry in two complementary senses. It preserves the relative positions of the DIV centroids, and it also gives a more faithful account of how the activity cloud expands and contracts during maturation.

\subsection{\textit{In silico} neuronal network response to stimulation}\label{sec:inscilicodriving}
As interest in biological machine learning continues to grow, understanding how neuronal networks respond to stimulation becomes increasingly important. A prevailing belief in the fields of learning and neuroplasticity is that appropriate stimulation strategies can guide network development toward desired structures, enabling the performance of specific tasks. Consequently, it is essential to evaluate the effectiveness of particular stimulation protocols to ensure that they successfully trigger learning. Without this validation, the potential for achieving meaningful outcomes in network training may be compromised.

\begin{figure*}
    \centering
    \begin{tabular}{@{}c@{\hspace{0.001\linewidth}}l@{\hspace{0.004\linewidth}}l@{\hspace{0.004\linewidth}}l@{}}
        & \makebox[0.302\linewidth][l]{A} & \makebox[0.302\linewidth][l]{B} & \makebox[0.302\linewidth][l]{C}\\[-0.20em]
        \raisebox{5.2em}[0pt][0pt]{\rotatebox{90}{\scriptsize weak stimulation}}
        & \includegraphics[width=0.302\linewidth, trim=0 0 0 0, clip]{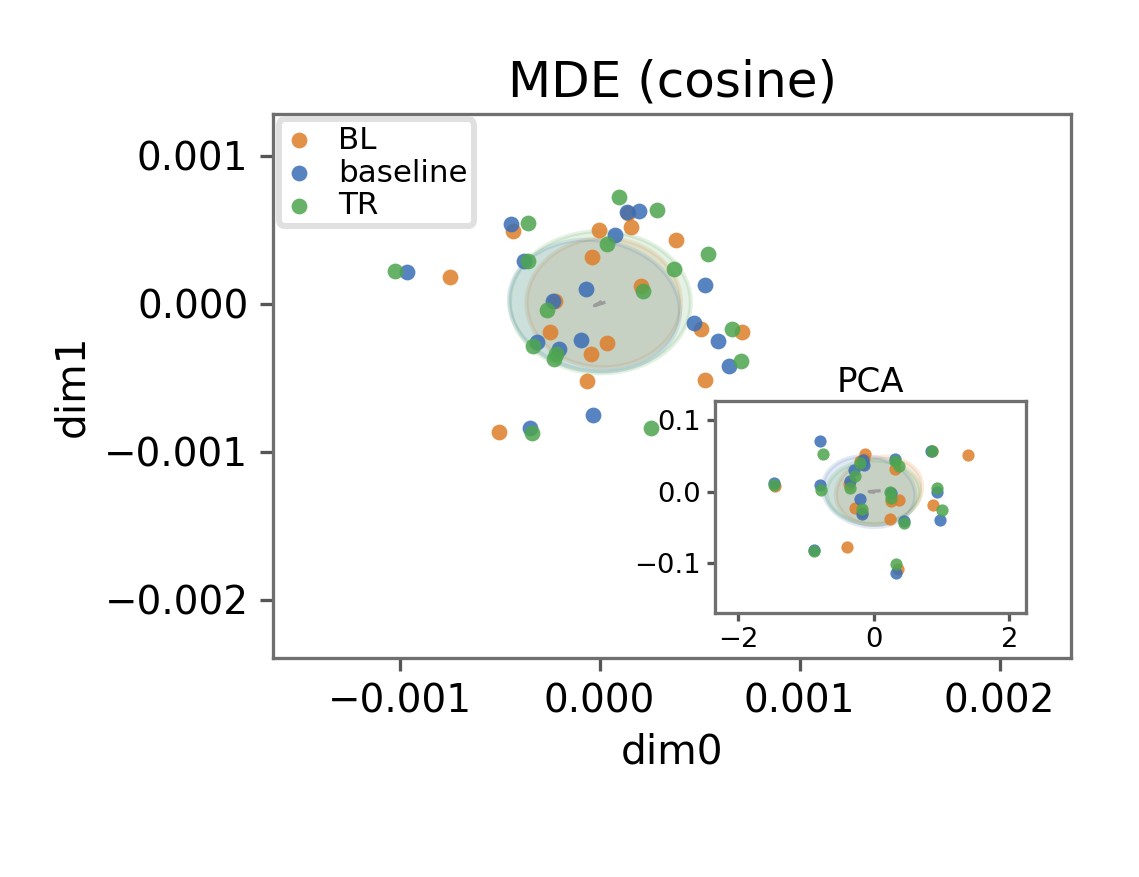}
        & \includegraphics[width=0.302\linewidth, trim=0 0 0 0, clip]{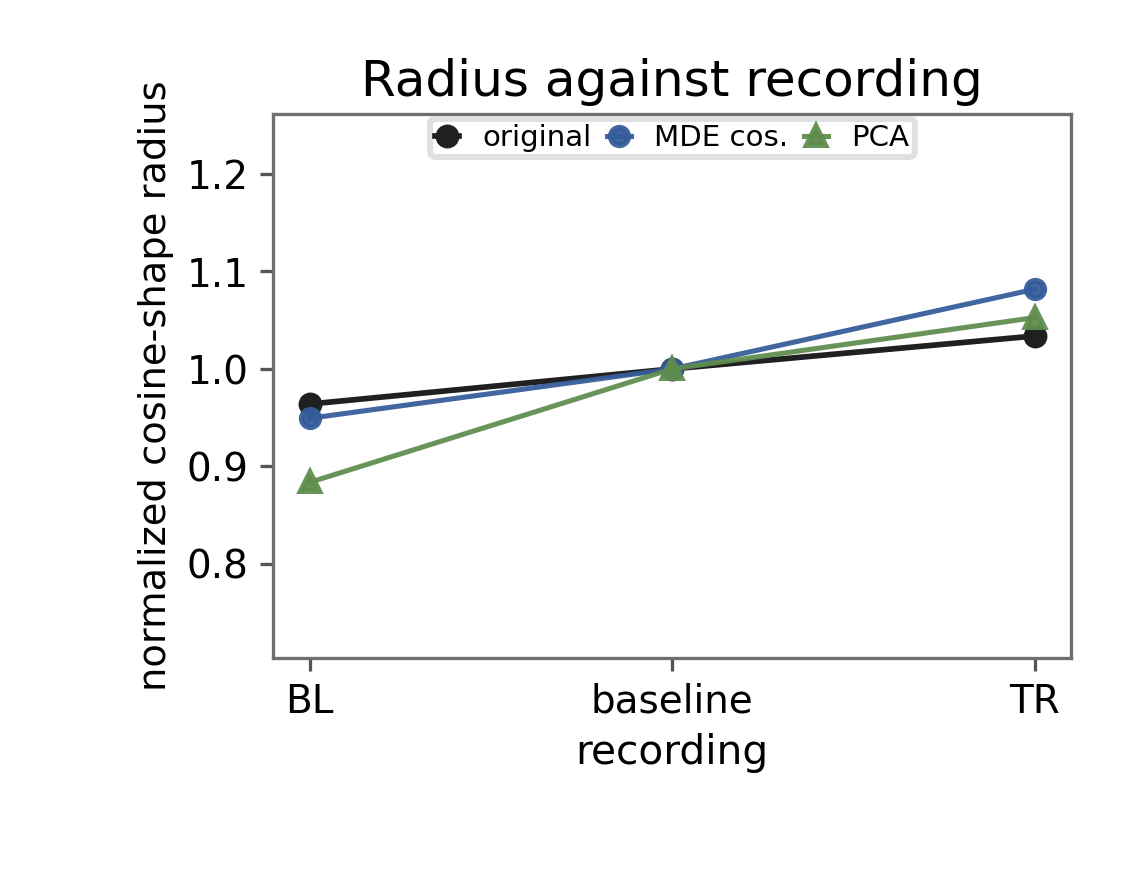}
        & \includegraphics[width=0.302\linewidth, trim=0 0 0 0, clip]{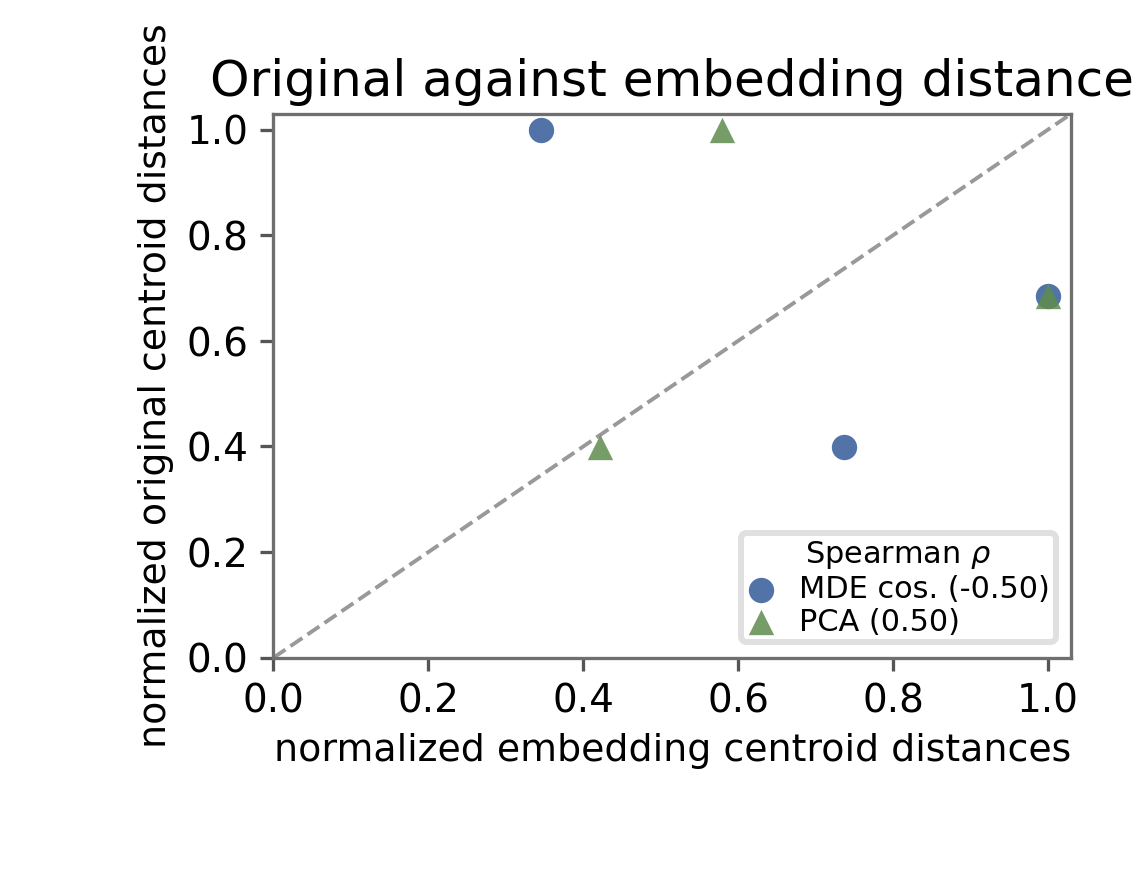}\\[-1.8em]
        & \makebox[0.302\linewidth][l]{D} & \makebox[0.302\linewidth][l]{E} & \makebox[0.302\linewidth][l]{F}\\[-0.20em]
        \raisebox{5.2em}[0pt][0pt]{\rotatebox{90}{\scriptsize strong stimulation}}
        & \includegraphics[width=0.302\linewidth, trim=0 0 0 0, clip]{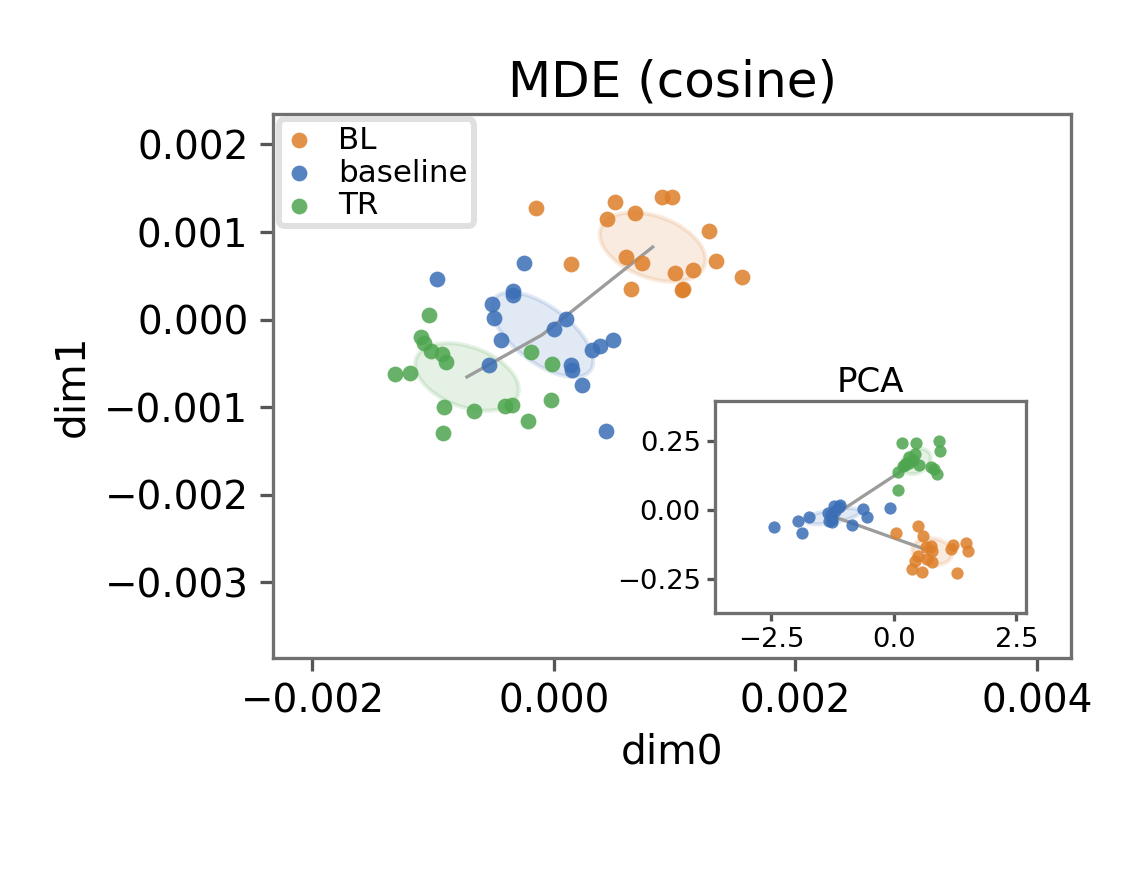}
        & \includegraphics[width=0.302\linewidth, trim=0 0 0 0, clip]{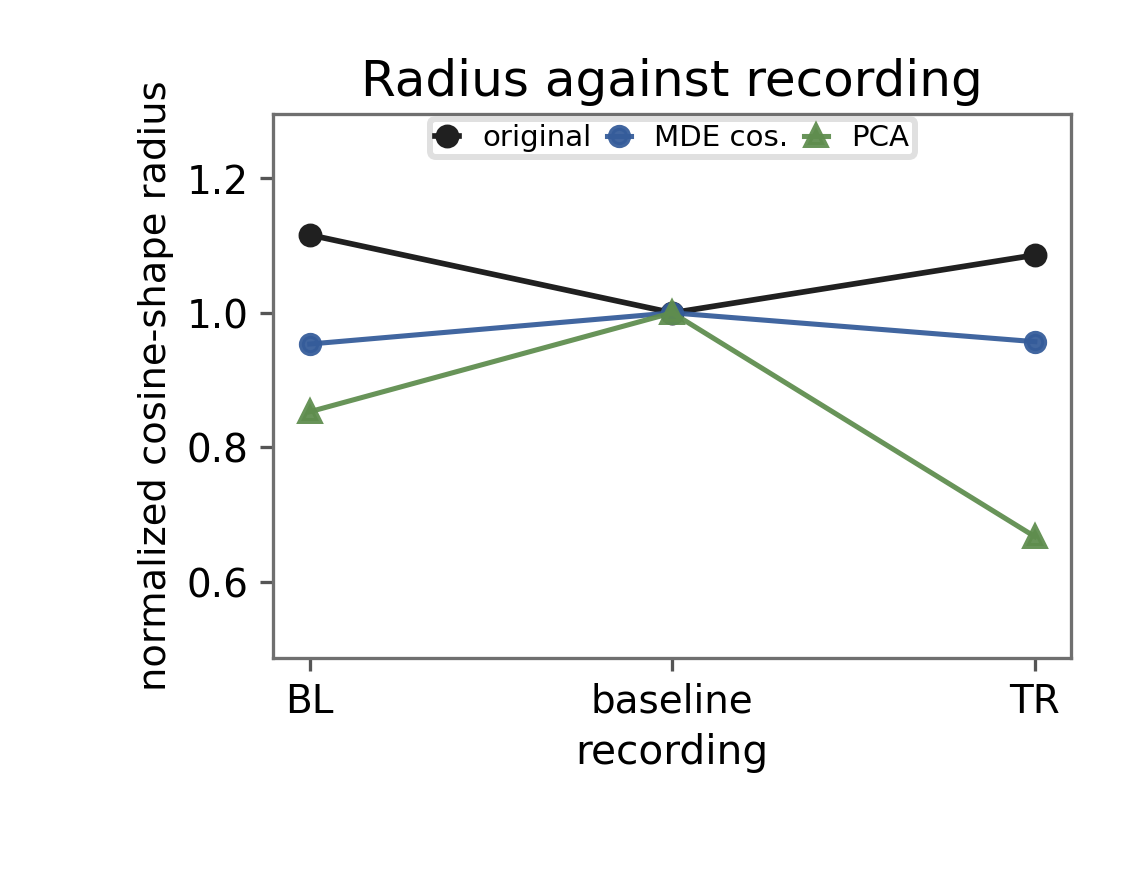}
        & \includegraphics[width=0.302\linewidth, trim=0 0 0 0, clip]{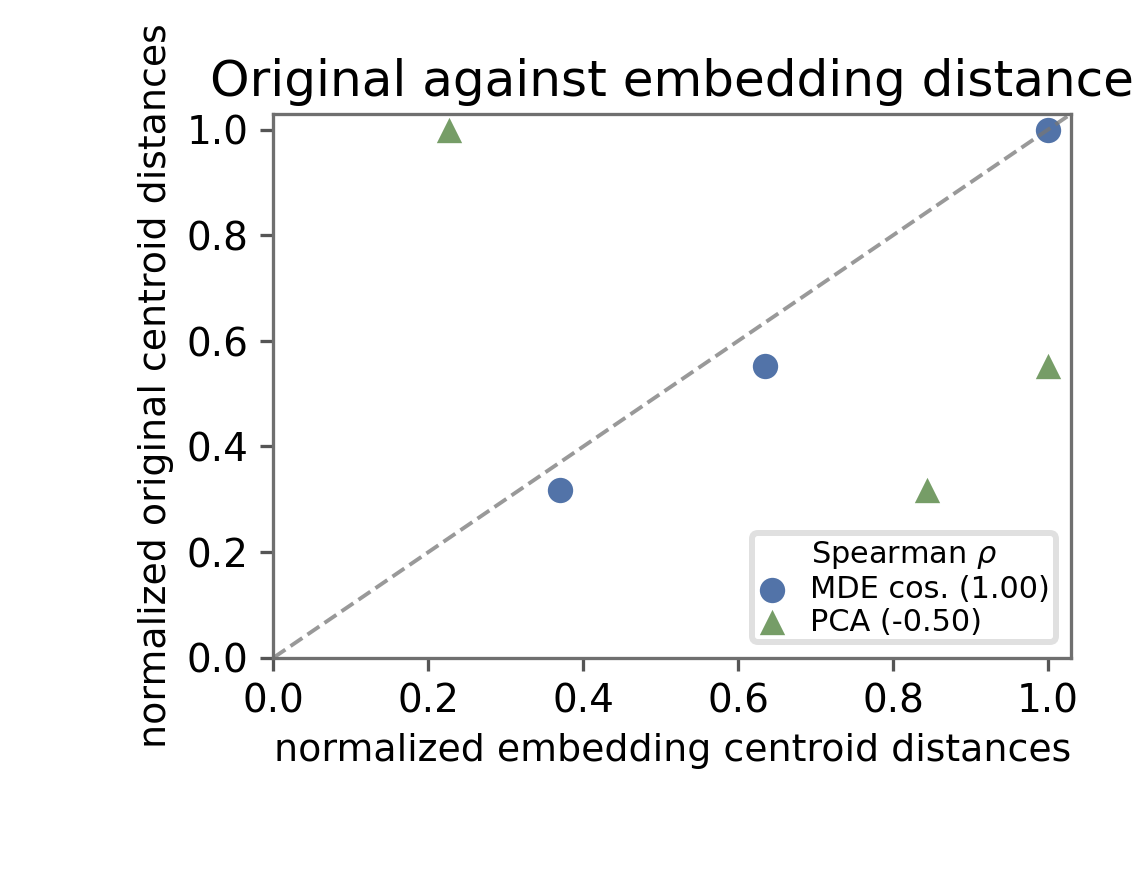}
    \end{tabular}
    \caption{Visualization and quantitative analysis of \emph{in silico} neuronal activity under weak and strong stimulation. A and D: MDE embeddings using a cosine metric, with PCA shown as an inset. Each dot corresponds to the analysis of a $50s$ window. The three recordings are baseline activity, stimulation to the bottom-left region (BL), and stimulation to the top-right region (TR). B and E: Normalized original cosine-shape radius and normalized embedding radius against recording, where all radii are normalized by the baseline recording. C and F: Normalized original centroid distances against normalized embedding centroid distances for all pairs of recordings, comparing MDE with a cosine metric and PCA.}
    \label{fig_silico_stim}
\end{figure*}

To study how visual informatics techniques can help identify the effectiveness of stimulation, we employ an \emph{in silico} model consisting of \(N=195\) neurons. We simulate two stimulation strengths and, for each strength, three recordings: baseline activity, stimulation applied to the bottom-left region (BL), and stimulation applied to the top-right region (TR). We compare MDE with a cosine metric against PCA, and evaluate the embeddings using the same quantities introduced above.

For weak stimulation, \fig{fig_silico_stim}A shows that the baseline, BL, and TR recordings strongly overlap in the MDE embedding. The PCA inset gives a similar picture, indicating that the weak stimulation does not move the population activity far away from the spontaneous state. This is also seen in \fig{fig_silico_stim}B. The original cosine-shape radius remains close to the baseline value, and both MDE and PCA reproduce the absence of a large expansion or contraction of the activity cloud. The centroid-distance diagnostic in \fig{fig_silico_stim}C is less stable, because the original centroid distances between the three recordings are very small. In this near-null regime, small changes in the embedding can change the rank ordering of the three pairwise distances.

For strong stimulation, the difference between the recordings becomes clearer. In \fig{fig_silico_stim}D, MDE separates the baseline, BL, and TR recordings, with the baseline activity located between the two stimulated conditions. PCA also separates the three recordings in the inset, so PCA is a useful qualitative baseline for this \emph{in silico} stimulation experiment. The quantitative diagnostics nevertheless show that MDE preserves the original geometry more faithfully. In \fig{fig_silico_stim}E, the original radius remains close to the baseline value under stimulation, while PCA gives a stronger contraction, especially for TR. MDE gives a radius closer to the original scale. More importantly, \fig{fig_silico_stim}F shows that MDE preserves the ordering of the pairwise centroid distances, whereas PCA does not. Thus, the advantage of MDE here is not that PCA completely fails to separate the stimulated conditions, but that MDE gives a more consistent two-dimensional representation with respect to the original population-activity geometry.

These results show why the \emph{in silico} stimulation experiment is useful. Weak stimulation provides a near-null case in which the activity remains close to the baseline state, while strong stimulation produces a clear displacement of the population activity. Visual embeddings can detect this difference, but the radius and centroid-distance diagnostics are needed to decide whether the displayed separation is geometrically faithful.

\subsection{\textit{In vitro} neuronal network in response to stimulation over time}
Following the \emph{in silico} stimulation experiment, we next apply the same analysis to an \emph{in vitro} rat cortical culture. We apply LTP stimulation to the network and study its activity before, during, and after stimulation. The recordings are divided into ten phases: 
Phase 0 corresponds to spontaneous activity; Phase 1 represents the activity during LTP stimulation; Phases 2-9 indicate the activity at 0, 10, 20, 30, 40, 50, 60 minutes, and 24 hours after stimulation, respectively. We compare MDE with a cosine metric against PCA, and again evaluate the embeddings using the original cosine-shape radius, the embedding radius, and the pairwise phase-centroid distances. This allows us to determine whether the two-dimensional map preserves the displacement of the activity state after stimulation, and whether it preserves the expansion or contraction of the activity cloud within each phase.

\begin{figure*}
    \centering
    \begin{tabular}{@{}c@{\hspace{0.015\linewidth}}c@{}}
        \begin{tabular}[t]{@{}l@{}}A\\[-0.30em]
        {\includegraphics[width=0.425\linewidth, trim=0 0 0 0, clip]{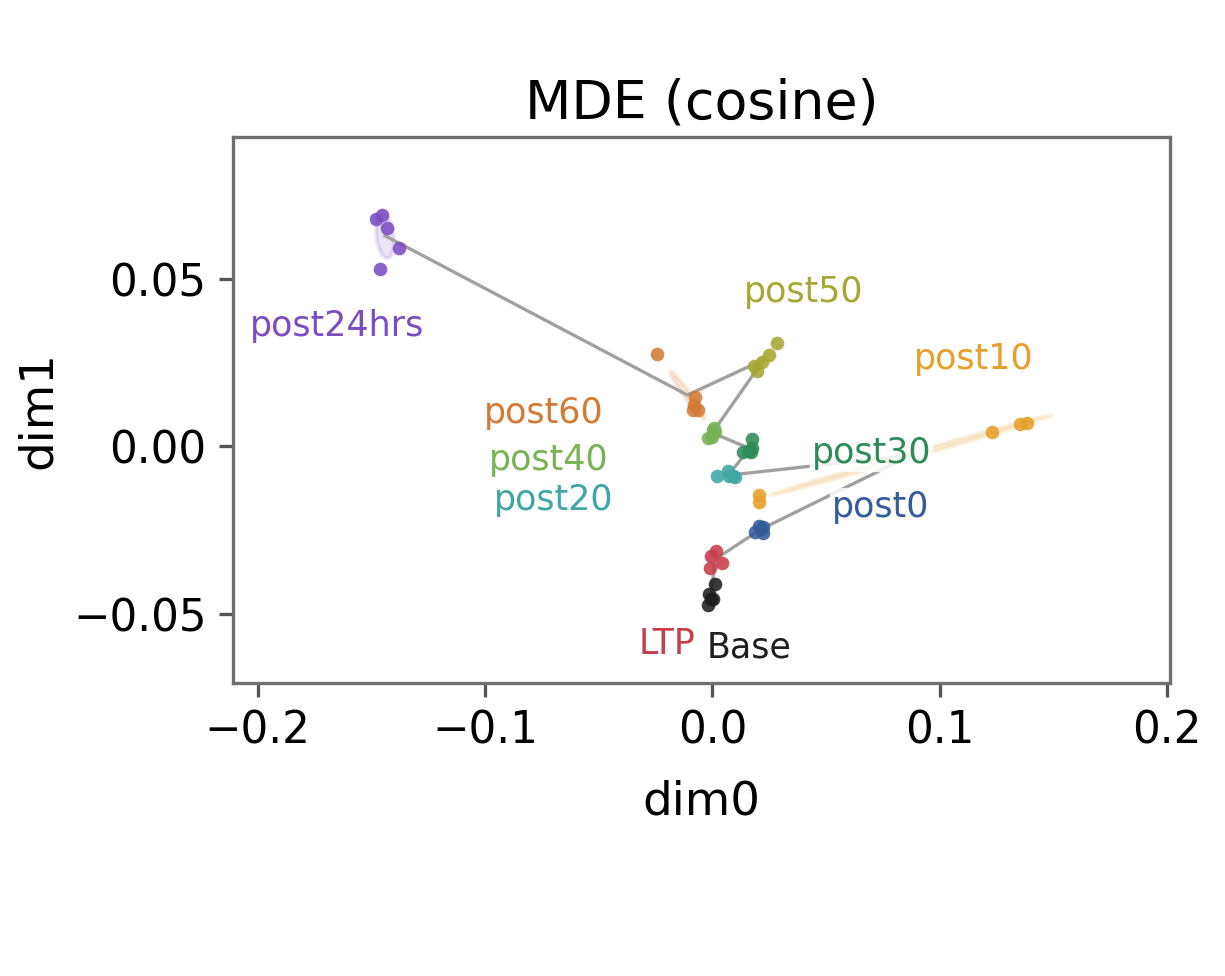}}\end{tabular}
        &
        \begin{tabular}[t]{@{}l@{}}B\\[-0.30em]
        {\includegraphics[width=0.425\linewidth, trim=0 0 0 0, clip]{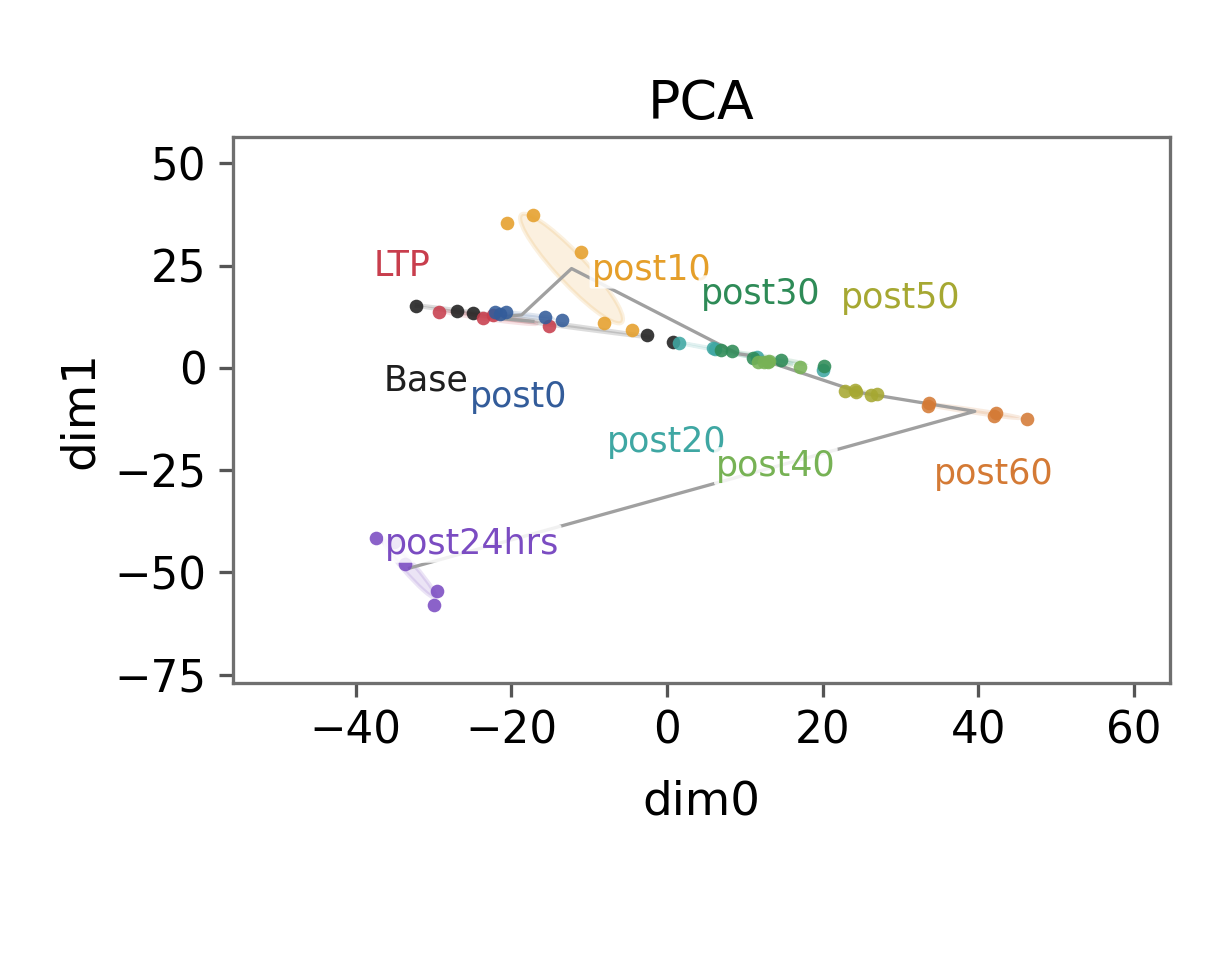}}\end{tabular}\\[-2.45em]
        \begin{tabular}[t]{@{}l@{}}C\\[-0.30em]
        {\includegraphics[width=0.425\linewidth, trim=0 0 0 0, clip]{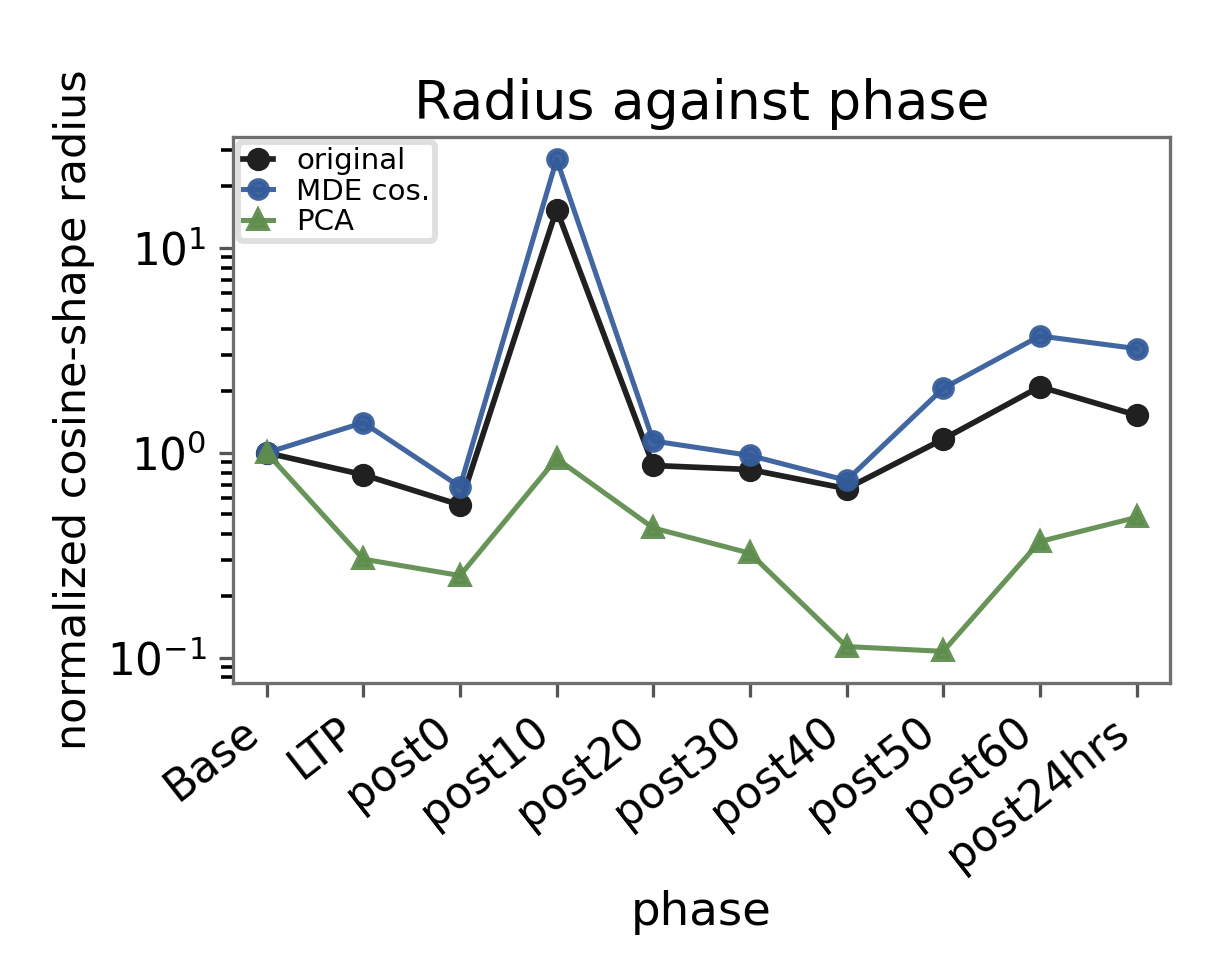}}\end{tabular}
        &
        \begin{tabular}[t]{@{}l@{}}D\\[-0.30em]
        {\includegraphics[width=0.425\linewidth, trim=0 0 0 0, clip]{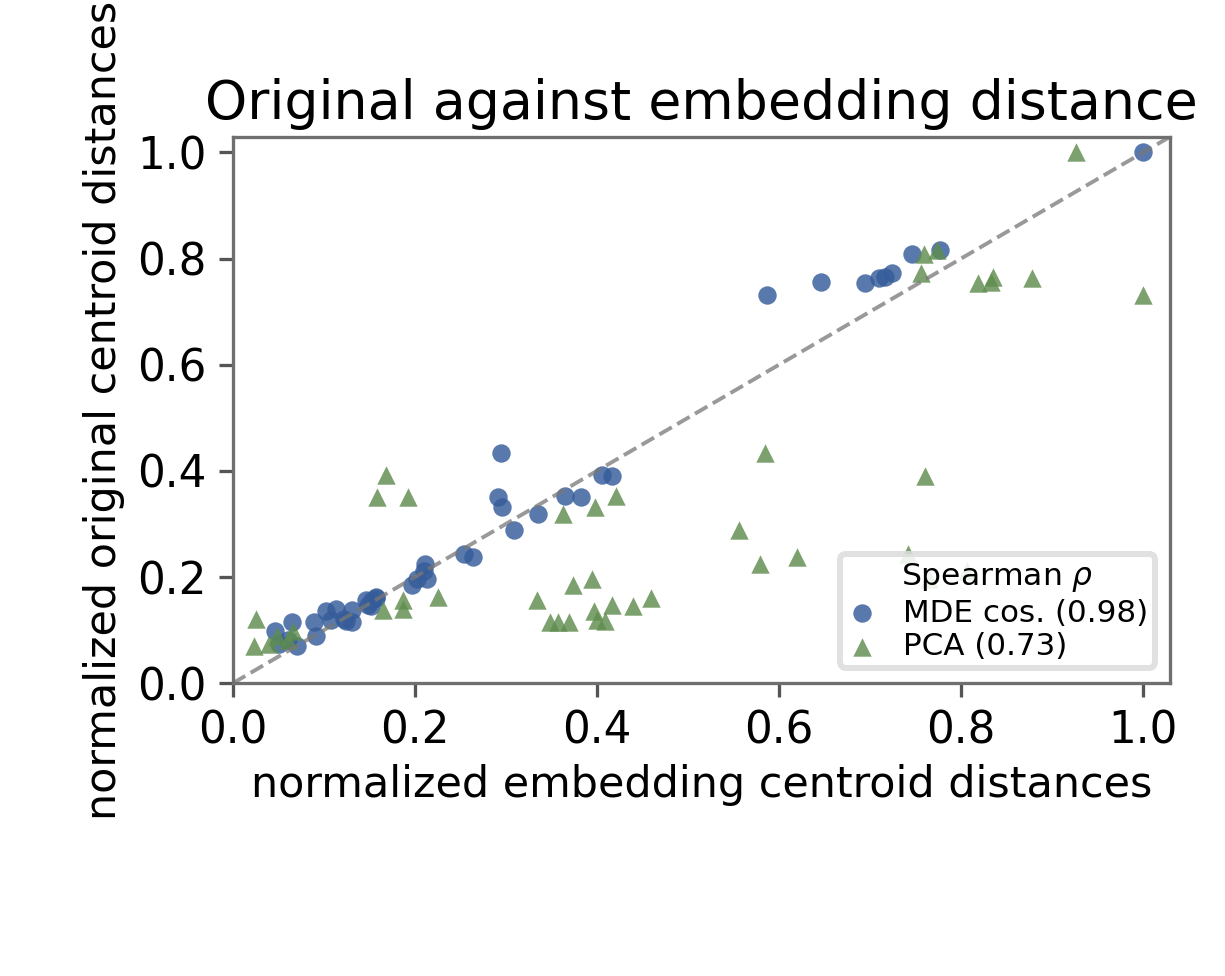}}\end{tabular}
    \end{tabular}
    \caption{Visualization and quantitative analysis of \emph{in vitro} rat cortical neuronal network activity before, during, and after LTP stimulation. A and B: MDE embedding using a cosine metric and PCA embedding, respectively. Each dot corresponds to the analysis of a $50s$ window, and colors indicate Base, LTP, post0, post10, post20, post30, post40, post50, post60, and post24hrs. C: Normalized original cosine-shape radius and normalized embedding radius against phase for the original activity space, MDE with a cosine metric, and PCA. A logarithmic y-axis is used because post10 gives a much larger radius than the other phases. D: Normalized original centroid distances against normalized embedding centroid distances for all pairs of phases, comparing MDE with a cosine metric and PCA.}
    \label{fig_in_vitro_stim}
\end{figure*}

The MDE embedding in \fig{fig_in_vitro_stim}A shows that the activity does not simply return to the spontaneous activity after LTP stimulation. Phase 0, Phase 1, and Phase 2 remain relatively close, whereas Phase 3, corresponding to 10 minutes after stimulation, is displaced to a different region of the embedding. The later post-stimulation phases then move through intermediate states, and Phase 9, corresponding to 24 hours after stimulation, is separated from the early phases. PCA in \fig{fig_in_vitro_stim}B also detects large changes in the activity, especially for Phase 9 and the later post-stimulation phases. However, the clusters are less clearly separated than in MDE, making the phase structure harder to interpret directly.

As shown in the cosine-shape radius of the original activity space in \fig{fig_in_vitro_stim}C, Phase 3 has a much larger cosine-shape radius than Phase 0, indicating that the population activity patterns become highly dispersed 10 minutes after stimulation. MDE with a cosine metric preserves this expansion, giving a large embedding radius at Phase 3. PCA does not capture this change, as its radius at Phase 3 remains close to the radius of the spontaneous activity, and several later phases are even compressed below the Phase 0 level.

As shown in \fig{fig_in_vitro_stim}D, MDE preserves the ordering of the original centroid distances very well, with a Spearman correlation of \(0.98\). PCA gives a lower correlation of \(0.73\), showing that it retains part of the large-scale separation but distorts several intermediate distances. Together, these results indicate that the LTP experiment contains two effects: the activity centroid moves through a sequence of post-stimulation states, with Phase 9 far from the early phases, and Phase 3 is a transient high-dispersion state. This is important because the separation between phase centroids and the spread within a phase are different quantities. A method can separate some phases visually while still missing the transient increase in within-phase variability.

In this section, we conducted four experiments utilizing both \emph{in silico} and \emph{in vitro} models to explore the dynamics of neuronal activity under varying stimulation conditions. For network development, MDE with a cosine metric preserves not only the visual trajectory of the conditions, but also the contraction or expansion of the activity cloud measured by the cosine-shape radius. This is seen in the \emph{in silico} maturation experiment and in the human cortical culture from DIV23 to DIV64. For stimulation, the weak \emph{in silico} experiment provides a near-null case, while strong stimulation and the \emph{in vitro} LTP experiment show clear displacement of the population activity. In the LTP experiment, MDE separates the phases more clearly than PCA and preserves the transient high-dispersion state at Phase 3, which PCA does not capture in the radius diagnostic. These experiments demonstrate the robustness of our methods across synthetic and biological systems, highlighting their potential for advancing our understanding of neuronal dynamics and the effects of stimulation. Further investigations into the underlying mechanisms of these interactions are planned and will be addressed in future work.

\section{Discussion}
The exploration of neuronal dynamics through high-dimensional spiking data is crucial for advancing our understanding of neuronal network development and the pathophysiology of neurological disorders. In this study, we showed an MDE-based visualization framework to analyze neuronal activity data. The main advantage of this framework is that it combines visual embeddings with quantitative diagnostics. Our findings demonstrate that the MDE-based approach summarizes key aspects of both geometric and over-time structure within complex neuronal datasets.

The results from our \emph{in silico} experiments highlight MDE's effectiveness in capturing the evolving connectivity patterns of simulated neuronal networks. MDE revealed a clear trajectory of network development as axon length increased, facilitating the visualization of the intricate dynamics of neuronal growth. The radius analysis further shows that this trajectory is accompanied by a contraction of the activity cloud as the simulated network matures. PCA can capture part of the dominant maturation direction, but it does not preserve this contraction. Furthermore, our \emph{in vitro} experiments further substantiate MDE's advantages, with distinct clustering of neuronal activity data observed from DIV23 to DIV64. In the human cortical culture, MDE gives a coherent developmental trajectory and preserves the relative DIV-centroid geometry better than PCA. This underscores MDE's ability to identify behavioral differences and connectivity changes as neuronal cultures mature, which is essential for understanding how networks adapt to intrinsic and extrinsic factors.

Our investigations into the effects of stimulation on neuronal activity yield pivotal insights into the plasticity and learning mechanisms of neuronal networks. The \emph{in silico} model demonstrates how to assess the effectiveness of stimulation strategies, which is vital for guiding network behavior. For weak stimulation, the activity remains close to the baseline state, whereas strong stimulation produces a clear displacement of the population activity. In the \emph{in vitro} LTP experiment, MDE forms clearer phase clusters than PCA and reveals that the activity after stimulation is not described by a simple return to the spontaneous state. The radius and centroid-distance diagnostics show that Phase 3 is a transient high-dispersion state, while Phase 9 is far from the early phases in centroid distance. This finding showcases MDE's applicability in exploring the dynamics of neuronal activity and the interplay between stimulation and intrinsic network properties.

The insights gained from this study have significant implications for future research in both basic and applied neuroscience, particularly in the context of cortical neuron-based machine learning. Identifying distinct phases in neuronal activity is crucial for developing effective cortical neuron-based computing devices. Understanding how networks change and adapt in response to stimulation can inform the design of biological computing systems that mimic learning and decision-making processes. The pairwise similarity graphs derived from the MDE can be used to phenotype neurons based on their temporal behavior, offering a more transparent (i.e., “white-box”) alternative to deep learning approaches such as \cite{pandarinath2018inferring}.

Another key contribution of this study is the emphasis on metric selection in dimensionality reduction techniques. Our results demonstrate that both MDE and t-SNE yield inadequate outcomes when employing Euclidean distance as a metric, as it fails to capture the complex, non-linear relationships in neuronal data. By utilizing cosine distance between population activity vectors, we achieved more meaningful embeddings, reinforcing the notion that careful consideration of distance metrics is essential for extracting valuable insights from high-dimensional datasets. This is especially important here because cosine distance compares the shape of population activity patterns, while the radius and centroid-distance diagnostics separately evaluate within-condition variability and between-condition displacement.

In summary, our research underscores the utility of advanced analytical techniques, such as MDE, in unraveling the complexities of neuronal dynamics. By preserving both global structures and local relationships, MDE provides a more comprehensive framework for understanding neuronal activity over time, particularly in the context of developmental processes and responses to stimulation. The results also show that a useful visualization should preserve both the relative positions of condition centroids and the size of each activity cloud. This distinction is important because an embedding may visually separate conditions while still missing changes in the underlying variability of the population activity. As we continue to explore the intricate workings of neuronal networks, the methodologies and findings presented in this study pave the way for future investigations aimed at enhancing our understanding of brain function and its implications for health and disease.

\section{Data Generation}

\subsection{\texorpdfstring{\emph{In silico} data generation}{In silico data generation}}
Neuronal culture activity is simulated using an \emph{in silico} neuronal culture model capturing the growth of network connections and the resulting neuronal activity~\cite{houben2025role}, tuned to qualitatively match the experimental data.
The model consists of two parts: the simulated growth of neurons' axons, in order to determine the network connectivity, and the simulation of the neuronal activity.
For the current study the neuronal activity simulation has been extended to capture the effect of external stimulation of the neurons on the neuronal activity.

Network growth is modeled following~\cite{orlandi2013noise}.
First, $N=195$ neurons are placed randomly on a two-dimensional circular area $3$~mm in diameter in a uniform manner, interpreting their somas as solid disks of radius $r = 7.5~\text{$\mu$m}$.
Following, axon growth is simulated by, starting from the center of each neuron, concatenating short line segments of length $\Delta\ell = 10~\text{$\mu$m}$, with each line segment $n$ placed with an angle $\varphi_n$ randomly deviating from that of the previous line segment: $\varphi_{n} = \varphi_{n-1} + \sigma_{\varphi}\mathcal{N}(0,1)$, resulting in quasi-linear axonal projections.
When the axon of a neuron $i$ reaches within a radius $r_j \sim \mathcal{N}(150, 20)~\text{$\mu$m}$ of the soma of neuron $j$, a connection from neuron $i$ to neuron $j$ is made with a $50\%$ probability.
The strength $w_{ji}$ of the connection is drawn from a Gaussian distribution with mean and width depending on the nature of the neuron $i$ (excitatory: $w_{ji} \sim \mathcal{N}(3, 0.6)$, inhibitory: $w_{ji} \sim \mathcal{N}(6, 1.2)$).
The total length of each neurons' axon $\ell_i$ is determined by drawing from a Rayleigh distribution with scale parameter $\sigma_L = \sqrt{2/\pi}L$, so that the mean axon length in the synthetic neuronal culture is $L~\text{mm}$.

Neuronal activity is modeled by spiking neurons coupled with dynamic synapses.
The neuron dynamics are simulated using the Izhikevich neuron model~\cite{izhikevich2003simple} in the regular spiking regime,
\begin{align*}
    \frac{dV_i}{dt} &= 0.04V_i^2 + 5 V_i + 140 - U_i + f_i(t) \\
    \frac{dU_i}{dt} &= 0.02(0.2V_i - U_i),
\end{align*}
where $V_i$ and $U_i$ (both dimensionless) are the membrane potential and recovery variable of the $i$-th neuron, respectively.
Each neuron is driven by
\begin{align*}
    f_i(t) = I_{i}(t) + \sigma \eta_i(t) + \sum_{j=1}^{N} w_{ij} P_j,
\end{align*}
which accounts for a combination of external stimulation $I_{i}$ described below, an independent Gaussian white-noise process $\eta_i(t)$ (with $\langle \eta_i(t) \eta_i(\tau) \rangle_t = \delta(t-\tau)$ and $\langle \eta_i(t) \eta_j(t) \rangle_t = \delta(i-j)$), and recurrent activity from the network consisting of a sum of pre-synaptic inputs $P_j(t)$ weighted by the connection weights $w_{ij}$. 

The pre-synaptic potential that each neuron induces on the neurons it projects to is determined by the synaptic dynamics, which simulate the post-synaptic potential trace $P_i$ and a synaptic vesicle reserve $R_i$,
\begin{align*}
    \frac{dP_i}{dt} &= -\frac{P_i}{\tau_P} + \sum_{k=1}^{n_i(t)} \int_{-\infty}^{t} R_i(t') \delta(t' - t_i^{(k)}) dt' \\
    \frac{dR_i}{dt} &= -\frac{1-R_i}{\tau_R} - \beta \sum_{k=1}^{n_i(t)} \int_{-\infty}^{t} R_i(t')\delta(t'-t_i^{(k)}) dt',
\end{align*}
where $n_i(t)$ is the total number of spikes emitted by neuron $i$ up to time $t$, and $t_i^{(k)}$ the time of the $k$-th spike of neuron $i$.
The constants $\tau_P = 10~\text{ms}$ and $\tau_R = 10\times 10^3~\text{ms}$ determine, respectively, the timescales of the decay of each post-synaptic potential $P_i$ and recovery of the synaptic vesicles $R_i$, and $\beta = 0.8$ determines the rate of synaptic depression due to vesicle depletion.

\subsubsection{Simulated development}
The changes in activity along development of the synthetic neuronal cultures are well captured by the increase of the average axon length $L$~\cite{montala2022rich, houben2025role}.
Hence, for the developmental dataset used in Sec.~\ref{sec:inscilicodevelopment} we have simulated axon growth for $L=1.5~\text{mm}$, but have taken snapshots of the connectivity along simulated growth at intervals of $0.05~\text{mm}$ in the average axon length, in other words: at $\langle \ell_i \rangle_i \in \{ 0.05, 0.1, 0.15, \ldots, 1.4, 1.45, 1.5 \}~\text{mm}$.
Following, for each snapshot $T = 30~\text{minutes}$ of activity is simulated.
In this dataset the external drive $I_i = 0$.

\subsubsection{Stimulation}
In section \ref{sec:inscilicodriving} we investigate the capacity of the visual informatics techniques to distinguish whether the neuronal networks are being stimulated or not.
To this end, electrical stimulation is introduced in the model neuronal culture by considering that it receives excitatory pulses of the form
\begin{align*}
    I_i(t) = g \sum_{k=1}^{m_i(t)} \int_{-\infty}^{t} \Theta{\left(t-t_i^{(k)}\right)} e^{-(t-t_i^{(k)})^2/\tau_I},
\end{align*}
which takes the form of exponentially decaying pulses with $m_i(t)$ the number of pulses received by neuron $i$ up to time $t$. $t_i^{(n)}$ denotes the time of the $k$-th input pulse, $g$ determines the amplitude of each pulse, and $\tau_I = 10~\text{ms}$ controls the decay of the post-synaptic effect of each pulse. 
The input effectively simulates additional excitatory pulses received by each neuron.

Specifically, for the data in sec.~\ref{sec:inscilicodriving} stimulation is carried out under two conditions: weak stimulation, with $g=2$, and strong stimulation, $g=4$.
Each condition consists of three phases: 1) unstimulated spontaneous activity; 2) stimulation to the neurons in the bottom-left quadrant of the culture; 3) stimulation to the neurons in the top-right quadrant of the culture.
The pulse times are generated following a homogeneous Poisson process with rate $\lambda = 0.1~\text{ms}^{-1}$ for each input train independently.

\subsection{\texorpdfstring{\emph{In vitro} data generation}{In vitro data generation}}
\subsubsection{Microelectrode array preparation and coating}

Neuronal cells were plated on a high-density MEAs chip (3Brain GmbH) containing 4096 electrodes ($64 \times 64$ grid) to monitor their activity. The chips were treated according to the commercially provided protocols of 3Brain. In detail, the chips were activated by loading their wells with 2 ml of dH$_2$O overnight to reduce their hydrophobic properties, followed by a sterilization with ethanol and UV light. Therefore, the wells were filled to the edges with 70\% ethanol. After 2 h of incubation, the wells were rinsed two times with sterile dH$_2$O. As a final sterilization step, the chips were cleaned with a tissue soaked in 70\% ethanol around the wells and placed for 15 min under UV light. To culture the cells, only the area of electrodes ($3.8 \times 3.8$~mm$^2$) was coated with poly-L-Ornithine (50 $\mu$g/ml, Sigma-Aldrich) and human laminin (20 $\mu$g/ml, BioLamina), each incubated overnight at 37~$^{\circ}$C. Before usage, the chips were rinsed two times with sterile dH$_2$O.

\subsubsection{\texorpdfstring{\emph{In vitro} rat primary cultures (excitatory and inhibitory)}{In vitro rat primary cultures (excitatory and inhibitory)}}

Embryonic rat cortices (E18, Charles River) were used to generate neuronal networks. Therefore, a mechanical dissection in ice-cold L-15 medium (Thermo Fisher Scientific), enriched with 0.6\% glucose and 0.5\% gentamicin (Sigma-Aldrich), was performed. First, the cortices were separated from the meninges. Secondly, they were mechanically dissociated by pipetting the tissue in Neurobasal Plus medium (Thermo Fisher Scientific) supplemented with Fetal Bovine Serum (10\%, Invitrogen), B27 plus (2\%, Thermo Fisher Scientific), GlutaMAX (1\%, Thermo Fisher Scientific), and Anti-Anti (0.1\%, Thermo Fisher Scientific). The final cell suspension was plated at a density of 2000 cells/mm$^2$ atop a pre-coated MEAs chip. The day of the dissection represents DIV 0. From here on, the cell cultures were maintained at 37~$^{\circ}$C, 5\% CO$_2$, and 95\% humidity. The cell medium was changed every other day. All procedures were approved by the Animal Experimentation Ethics Committee (CEEA) of the University of Barcelona, under order DMAH-5461, in accordance with the regulations of the Generalitat de Catalunya (Spain).

\subsubsection{\texorpdfstring{\emph{In vitro} human iPSC-derived cortical cultures (purely excitatory)}{In vitro human iPSC-derived cortical cultures (purely excitatory)}}

Purely excitatory human-induced neurons previously characterized by~\cite{Parodi_2023} were used. To obtain human-induced neurons, the protocol of~\cite{WANG2023101967} was followed, in which rtTA/Ngn2-positive human-iPSC~\cite{frega2017rapid} were differentiated into neurons due to the transcription factor neurogenin-2. In detail, the frozen supplied human iPSCs were cultured in E8Flex medium (Thermo Fisher Scientific) supplemented with E8F supplement (2\%, Thermo Fisher Scientific), G418 (50 $\mu$g/ml, Sigma-Aldrich), pen/streptomycin (1\%, Sigma-Aldrich), and puromycin (0.5 $\mu$g/ml, Sigma-Aldrich). After thawing or splitting the cells, the medium was enriched with RevitaCell (1\%, Thermo Fisher Scientific). Every other day, half of the cell medium was changed. To start the differentiation of the iPSCs into neurons, the cell medium was supplemented with doxycycline (4 $\mu$g/ml, Sigma-Aldrich), leading to the expression of neurogenin-2 and thus represented day DIV 0 of the differentiation. At DIV 3, cells were detached via Accutase (Sigma-Aldrich) and co-cultured (70:30 ratio) with rat cortical astrocytes (Thermo Fisher Scientific) on pre-cleaned and coated MEA chips. A total cell concentration of 2000 cells/mm$^2$ was plated on top of the electrodes. From here on, cells were cultured in Neurobasal Plus medium (Thermo Fisher Scientific) supplemented with B27 plus supplement (2\%, Thermo Fisher Scientific), GlutaMAX (1\%, Thermo Fisher Scientific), penicillin/streptomycin (1\%, Sigma-Aldrich), Human Neurotrophin-3 (NT-3, 10 ng/ml, Sigma-Aldrich), doxycycline (4 $\mu$g/ml, Sigma-Aldrich), and Human Brain-Derived Neurotrophic Factor (BDNF, 10 ng/ml, Sigma-Aldrich). At DIV 16, the cell medium was enriched with 2\% Fetal Bovine Serum (Thermo Fisher Scientific), and at DIV 23, doxycycline was removed from the medium. Cultures were incubated at 37$^{\circ}$C, 5\% CO$_2$, and 95\% humidity, and half of the medium was replaced every other day.

\subsubsection{MEAa recordings and stimulations}

Neuronal activity was acquired with the BioCam DupleX system (3Brain GmbH) at a sampling rate of 20~kHz. All recordings and stimulation phases were performed at 34~$^{\circ}$C and 5\% CO$_2$. The neuronal development was tracked by recording the activity throughout differentiation days, starting early at 23 DIV up to later stages of maturation of 64 DIV, each recording lasting 10 min.

To stimulate the cells, all electrodes could be chosen to send an electrical current from one electrode to another. In this setup, the electrical current was sent from a selected electrode to the left neighbor electrode. For the stimulation, 15 electrode pairs with a random location on the chip were selected. The only selection criteria were that the electrodes displayed neuronal activity during a collective burst event and that the electrodes responded to a single pulse stimulation. In this stimulation setup, tetanic stimulation was performed ~\cite{Chiappalone08}. In detail, 10 bipolar pulses (+/-) of 10 $\mu$Amp and 100 $\mu$s were released at 20 Hz. The stimulation train was repeated 20 times at 0.1 Hz. Different time points before, during and after stimulation were recorded (5 min each), as follows:

\begin{enumerate}
    \setlength{\itemsep}{-2pt} 
    \item[] Phase 0: Spontaneous activity.
    \item[] Phase 1: Tetanic stimulation.
    \item[] Phase 2: Just after stimulation.
    \item[] Phase 3: 10 min post-stimulation.
    \item[] Phase 4: 20 min post-stimulation.
    \item[] Phase 5: 30 min post-stimulation.
    \item[] Phase 6: 40 min post-stimulation.
    \item[] Phase 7: 50 min post-stimulation.
    \item[] Phase 8: 60 min post-stimulation.
    \item[] Phase 9: 24 h post-stimulation.
\end{enumerate}
\vspace{-10pt}

\subsubsection{Spike detection}

To obtain the spikes, the BrainWave5 software was used. Therefore, the raw data was filtered through a high-pass filter of 800 Hz, and a Precise Timing Spike Detection (PTSD) was performed \cite{MACCIONE2009241}, with parameters provided in Table 1.

\begin{table}[h]
    \centering
    \begin{tabular}{|l|l|}
        \hline
        Standard deviation factor & 8 \\
        Peak lifetime period & 2 ms \\
        Spike assignment & Fast varying peak \\
        AI-validation & Oc \\
        Pre-peak duration & 1 ms \\
        Post-peak duration & 1.5 ms \\
        Discard noisy electrodes & On \\
        Discard chip calibration artefacts & On \\
        \hline
    \end{tabular}
    \caption{Parameters of BrainWave5 software for PTSD spike detection.}
\end{table}

\subsection{Ethical approvals}

All procedures containin the dissection of the\textit{ in vitro} rat embryonic cortices and preparation of primary neuronal cultures  were approved by the Animal Experimentation Ethics Committee (CEEA) of the University of Barcelona, under order DMAH-5461, in accordance with the regulations of the Generalitat de Catalunya (Spain).

\subsection{Data and code availability}

All data presented in this paper is available from \url{https://doi.org/10.17036/researchdata.aston.ac.uk.00000654}. This includes the Python file containing the derived algorithm, as well as both the in-silico and in-vitro neuronal firing data being studied.

\section*{Acknowledgements}
This research is supported by the European Union Horizon 2020 research and innovation program under Grant No. 964977 (project NEU-CHiP). HFP and DS are supported from the UK Multidisciplinary Centre for Neuromorphic Computing (UKRI982) is gratefully acknowledged. JS also acknowledges financial support from the Spanish Ministerio de Ciencia e Innovación under project PID2022-137713NB-C22 and by the Generalitat de Catalunya under project 2021-SGR-00450. DT also acknowledges the support from Ministerio de Ciencia e Innovación (Spain), under projects No. CNS2023-143862. The authors acknowledge the support of Aston University Biomedical Facility for the purpose of providing infrastructure support within the College of Health and Life Sciences. Further, the authors would like to thank G. Parodi from S. Martinoia's group for the support in the preparation of human-induced neuronal networks, as well N.N. Kasri for the supply of the human-iPSC cell line.

\section*{AUTHOR CONTRIBUTIONS}
H.F.P., Y.P.R and D.S. designed research, derived algorithm, and analyzed all data. A.C.H. performed \emph{in vitro} experiments; A.M.H. performed \emph{in silico} experiments. H.F.P., A.M.H., A.C.H., Y.P.R, D.T., J.S. and D.S. wrote the paper.

\bibliographystyle{prsty}

\bibliography{visual_ref}

@article{maccione2010experimental,
	
	AUTHOR={Maccione, Alessandro and Gandolfo, Mauro and Tedesco, MariaTeresa and Nieus, Thierry and Imfeld, Kilian and Martinoia, Sergio and Luca, Berdondini},   
	
	TITLE={Experimental investigation on spontaneously active hippocampal cultures recorded by means of high-density MEAs: analysis of the spatial resolution effects},      
	
	JOURNAL={Frontiers in Neuroengineering},      
	
	VOLUME={3},           
	
	YEAR={2010},      

	
	ISSN={1662-6443},   
}

@article{pandarinath2018inferring,
  title={Inferring single-trial neural population dynamics using sequential auto-encoders},
  author={Pandarinath, Chethan and O’Shea, Daniel J and Collins, Jasmine and Jozefowicz, Rafal and Stavisky, Sergey D and Kao, Jonathan C and Trautmann, Eric M and Kaufman, Matthew T and Ryu, Stephen I and Hochberg, Leigh R and others},
  journal={Nature methods},
  volume={15},
  number={10},
  pages={805--815},
  year={2018},
  publisher={Nature Publishing Group US New York}
}

@article{mcinnes2018umap, doi = {10.21105/joss.00861}, url = {https://doi.org/10.21105/joss.00861}, year = {2018}, publisher = {The Open Journal}, volume = {3}, number = {29}, pages = {861}, author = {McInnes, Leland and Healy, John and Saul, Nathaniel and Großberger, Lukas}, title = {UMAP: Uniform Manifold Approximation and Projection}, journal = {Journal of Open Source Software} }

@article{cunningham2014dimensionality,
  title={Dimensionality reduction for large-scale neural recordings},
  author={Cunningham, John P and Yu, Byron M},
  journal={Nature neuroscience},
  volume={17},
  number={11},
  pages={1500--1509},
  year={2014},
  publisher={Nature Publishing Group US New York}
}

@inproceedings{lawrence2003gaussian,
author = {Lawrence, Neil D.},
title = {Gaussian process latent variable models for visualisation of high dimensional data},
year = {2003},
publisher = {MIT Press},
address = {Cambridge, MA, USA},
booktitle = {Proceedings of the 17th International Conference on Neural Information Processing Systems},
pages = {329–336},
numpages = {8},
location = {Whistler, British Columbia, Canada},
series = {NIPS'03}
}

@book{kim1978introduction,
	author = {Kim, Jae-On and Mueller, Charles W.},
	title = {Introduction to Factor Analysis : What it is and how to do it},
	publisher = {SAGE},
	year = {1978},
	series = {Quantitative Applications in the Social Sciences},
	address = {New Delhi}
}

@InProceedings{scholkopf1997kernel,
author="Sch{\"o}lkopf, Bernhard
and Smola, Alexander
and M{\"u}ller, Klaus-Robert",
editor="Gerstner, Wulfram
and Germond, Alain
and Hasler, Martin
and Nicoud, Jean-Daniel",
title="Kernel principal component analysis",
booktitle="Artificial Neural Networks --- ICANN'97",
year="1997",
publisher="Springer Berlin Heidelberg",
address="Berlin, Heidelberg",
pages="583--588",
isbn="978-3-540-69620-9"
}

@article{farooq2024adaptive,
  author  = {Adam Farooq and Yordan P. Raykov and Petar Raykov and Max A. Little},
  title   = {Adaptive Latent Feature Sharing for Piecewise Linear Dimensionality Reduction},
  journal = {Journal of Machine Learning Research},
  year    = {2024},
  volume  = {25},
  number  = {135},
  pages   = {1--42}
}

@article{agrawal2021minimum,
  title={Minimum-distortion embedding},
  author={Agrawal, Akshay and Ali, Alnur and Boyd, Stephen and others},
  journal={Foundations and Trends{\textregistered} in Machine Learning},
  volume={14},
  number={3},
  pages={211--378},
  year={2021},
  publisher={Now Publishers, Inc.}
}

@article{van2008visualizing,
  title={Visualizing data using t-SNE.},
  author={Van der Maaten, Laurens and Hinton, Geoffrey},
  journal={Journal of machine learning research},
  volume={9},
  number={11},
  year={2008}
}

@article{pearson1901liii,
  title={LIII. On lines and planes of closest fit to systems of points in space},
  author={Pearson, Karl},
  journal={The London, Edinburgh, and Dublin philosophical magazine and journal of science},
  volume={2},
  number={11},
  pages={559--572},
  year={1901},
  publisher={Taylor \& Francis}
}

@article{grienberger2012imaging,
  title={Imaging calcium in neurons},
  author={Grienberger, Christine and Konnerth, Arthur},
  journal={Neuron},
  volume={73},
  number={5},
  pages={862--885},
  year={2012},
  publisher={Elsevier}
}

@article{kim2022fluorescence,
  title={Fluorescence imaging of large-scale neural ensemble dynamics},
  author={Kim, Tony Hyun and Schnitzer, Mark J},
  journal={Cell},
  volume={185},
  number={1},
  pages={9--41},
  year={2022},
  publisher={Elsevier}
}

@inproceedings{strubell2019energy,
	title = "Energy and Policy Considerations for Deep Learning in {NLP}",
	author = "Strubell, Emma  and
	Ganesh, Ananya  and
	McCallum, Andrew",
	editor = "Korhonen, Anna  and
	Traum, David  and
	M{\`a}rquez, Llu{\'\i}s",
	booktitle = "Proceedings of the 57th Annual Meeting of the Association for Computational Linguistics",
	month = jul,
	year = "2019",
	address = "Florence, Italy",
	publisher = "Association for Computational Linguistics",
	pages = "3645--3650",
}

@article{cai2023brain,
  title={Brain organoid reservoir computing for artificial intelligence},
  author={Cai, Hongwei and Ao, Zheng and Tian, Chunhui and Wu, Zhuhao and Liu, Hongcheng and Tchieu, Jason and Gu, Mingxia and Mackie, Ken and Guo, Feng},
  journal={Nature Electronics},
  volume={6},
  number={12},
  pages={1032--1039},
  year={2023},
  publisher={Nature Publishing Group UK London}
}

@article{kagan2022vitro,
  title={In vitro neurons learn and exhibit sentience when embodied in a simulated game-world},
  author={Kagan, Brett J and Kitchen, Andy C and Tran, Nhi T and Habibollahi, Forough and Khajehnejad, Moein and Parker, Bradyn J and Bhat, Anjali and Rollo, Ben and Razi, Adeel and Friston, Karl J},
  journal={Neuron},
  volume={110},
  number={23},
  pages={3952--3969},
  year={2022},
  publisher={Elsevier}
}

@article{sotirakis2023identification,
  title={Identification of motor progression in Parkinson’s disease using wearable sensors and machine learning},
  author={Sotirakis, Charalampos and Su, Zi and Brzezicki, Maksymilian A and Conway, Niall and Tarassenko, Lionel and FitzGerald, James J and Antoniades, Chrystalina A},
  journal={npj Parkinson's Disease},
  volume={9},
  number={1},
  pages={142},
  year={2023},
  publisher={Nature Publishing Group UK London}
}

@article{iosa2022principal,
  title={Principal component analysis of oxford cognitive screen in patients with stroke},
  author={Iosa, Marco and Demeyere, Nele and Abbruzzese, Laura and Zoccolotti, Pierluigi and Mancuso, Mauro},
  journal={Frontiers in neurology},
  volume={13},
  pages={779679},
  year={2022},
  publisher={Frontiers Media SA}
}

@article{daddinounou2024bi,
  title={Bi-sigmoid spike-timing dependent plasticity learning rule for magnetic tunnel junction-based SNN},
  author={Daddinounou, Salah and Vatajelu, Elena-Ioana},
  journal={Frontiers in Neuroscience},
  volume={18},
  pages={1387339},
  year={2024},
  publisher={Frontiers Media SA}
}

@article{pedretti2017memristive,
  title={Memristive neural network for on-line learning and tracking with brain-inspired spike timing dependent plasticity},
  author={Pedretti, G and Milo, V and Ambrogio, S and Carboni, R and Bianchi, S and Calderoni, A and Ramaswamy, N and Spinelli, AS and Ielmini, D},
  journal={Scientific reports},
  volume={7},
  number={1},
  pages={5288},
  year={2017},
  publisher={Nature Publishing Group UK London}
}

@article{li2023short,
  title={Short-term synaptic plasticity in emerging devices for neuromorphic computing},
  author={Li, Chao and Zhang, Xumeng and Chen, Pei and Zhou, Keji and Yu, Jie and Wu, Guangjian and Xiang, Du and Jiang, Hao and Wang, Ming and Liu, Qi},
  journal={Iscience},
  volume={26},
  number={4},
  year={2023},
  publisher={Elsevier}
}

@article{badre2021dimensionality,
  title={The dimensionality of neural representations for control},
  author={Badre, David and Bhandari, Apoorva and Keglovits, Haley and Kikumoto, Atsushi},
  journal={Current Opinion in Behavioral Sciences},
  volume={38},
  pages={20--28},
  year={2021},
  publisher={Elsevier}
}

@article{mccready2022multielectrode,
  title={Multielectrode arrays for functional phenotyping of neurons from induced pluripotent stem cell models of neurodevelopmental disorders},
  author={McCready, Fraser P and Gordillo-Sampedro, Sara and Pradeepan, Kartik and Martinez-Trujillo, Julio and Ellis, James},
  journal={Biology},
  volume={11},
  number={2},
  pages={316},
  year={2022},
  publisher={MDPI}
}

@article{chung2021neural,
  title={Neural population geometry: An approach for understanding biological and artificial neural networks},
  author={Chung, SueYeon and Abbott, Larry F},
  journal={Current opinion in neurobiology},
  volume={70},
  pages={137--144},
  year={2021},
  publisher={Elsevier}
}

@article{pang2016dimensionality,
  title={Dimensionality reduction in neuroscience},
  author={Pang, Rich and Lansdell, Benjamin J and Fairhall, Adrienne L},
  journal={Current Biology},
  volume={26},
  number={14},
  pages={R656--R660},
  year={2016},
  publisher={Elsevier}
}

@article{shinn2023phantom,
  title={Phantom oscillations in principal component analysis},
  author={Shinn, Maxwell},
  journal={Proceedings of the National Academy of Sciences},
  volume={120},
  number={48},
  pages={e2311420120},
  year={2023},
  publisher={National Acad Sciences}
}

@article{kobak2016demixed,
  title={Demixed principal component analysis of neural population data},
  author={Kobak, Dmitry and Brendel, Wieland and Constantinidis, Christos and Feierstein, Claudia E and Kepecs, Adam and Mainen, Zachary F and Qi, Xue-Lian and Romo, Ranulfo and Uchida, Naoshige and Machens, Christian K},
  journal={elife},
  volume={5},
  pages={e10989},
  year={2016},
  publisher={eLife Sciences Publications, Ltd}
}

@article{zhou2020visualization,
  title={Visualization of single cell RNA-seq data using t-SNE in R},
  author={Zhou, Bo and Jin, Wenfei},
  journal={Stem Cell Transcriptional Networks: Methods and Protocols},
  pages={159--167},
  year={2020},
  publisher={Springer}
}

@article{hu2020t,
  title={T-distribution stochastic neighbor embedding for fine brain functional parcellation on rs-fMRI},
  author={Hu, Ying and Li, Xiaobing and Wang, Lijia and Han, Baosan and Nie, Shengdong},
  journal={Brain Research Bulletin},
  volume={162},
  pages={199--207},
  year={2020},
  publisher={Elsevier}
}

@article{kobak2019art,
  title={The art of using t-SNE for single-cell transcriptomics},
  author={Kobak, Dmitry and Berens, Philipp},
  journal={Nature communications},
  volume={10},
  number={1},
  pages={5416},
  year={2019},
  publisher={Nature Publishing Group UK London}
}

@inproceedings{badoiu2005low,
author = {Badoiu, Mihai and Chuzhoy, Julia and Indyk, Piotr and Sidiropoulos, Anastasios},
title = {Low-distortion embeddings of general metrics into the line},
year = {2005},
isbn = {1581139608},
publisher = {Association for Computing Machinery},
address = {New York, NY, USA},
booktitle = {Proceedings of the Thirty-Seventh Annual ACM Symposium on Theory of Computing},
pages = {225--233},
numpages = {9},
location = {Baltimore, MD, USA},
series = {STOC '05}
}

@article{fomin2011exact,
  title={An exact algorithm for minimum distortion embedding},
  author={Fomin, Fedor V and Lokshtanov, Daniel and Saurabh, Saket},
  journal={Theoretical computer science},
  volume={412},
  number={29},
  pages={3530--3536},
  year={2011},
  publisher={Elsevier}
}

@article{zhu2015different,
  title={Different patterns of electrical activity lead to long-term potentiation by activating different intracellular pathways},
  author={Zhu, Guoqi and Liu, Yan and Wang, Yubin and Bi, Xiaoning and Baudry, Michel},
  journal={Journal of Neuroscience},
  volume={35},
  number={2},
  pages={621--633},
  year={2015},
  publisher={Soc Neuroscience}
}

@article{escobar2024long,
  title={Long-term bilateral change in pain and sensitivity to high-frequency cutaneous electrical stimulation in healthy subjects depends on stimulus modality: a dermatomal examination},
  author={Escobar-S{\'a}nchez, Isabel and R{\'\i}os-Le{\'o}n, Marta and Taylor, Julian},
  journal={Frontiers in Medicine},
  volume={10},
  pages={1337711},
  year={2024},
  publisher={Frontiers Media SA}
}

@article{houben2025role,
  title={Role of connectivity anisotropies in the dynamics of cultured neuronal networks},
  author={Houben, Akke Mats and Garcia-Ojalvo, Jordi and Soriano, Jordi},
  journal={PLOS Computational Biology},
  volume={21},
  number={11},
  pages={e1012727},
  year={2025},
  publisher={Public Library of Science San Francisco, CA USA}
}

@article{orlandi2013noise,
  title={Noise focusing and the emergence of coherent activity in neuronal cultures},
  author={Orlandi, Javier G and Soriano, Jordi and Alvarez-Lacalle, Enrique and Teller, Sara and Casademunt, Jaume},
  journal={Nature Physics},
  volume={9},
  number={9},
  pages={582--590},
  year={2013},
  publisher={Nature Publishing Group UK London}
}

@article{izhikevich2003simple,
	author={Izhikevich, E.M.},
	journal={IEEE Transactions on Neural Networks}, 
	title={Simple model of spiking neurons}, 
	year={2003},
	volume={14},
	number={6},
	pages={1569-1572}}

@article{montala2022rich,
    title = {Rich dynamics and functional organization on topographically designed neuronal networks in vitro},
    journal = {iScience},
    volume = {25},
    number = {12},
    pages = {105680},
    year = {2022},
    issn = {2589-0042},
    author = {Marc Montalà-Flaquer and Clara F. López-León and Daniel Tornero and Akke Mats Houben and Tanguy Fardet and Pascal Monceau and Samuel Bottani and Jordi Soriano}
}

@article{Parodi_2023,
year = {2023},
month = {sep},
publisher = {IOP Publishing},
volume = {20},
number = {5},
pages = {056011},
author = {Parodi, Giulia and Brofiga, Martina and Pastore, Vito Paolo and Chiappalone, Michela and Martinoia, Sergio},
title = {Deepening the role of excitation/inhibition balance in human iPSCs-derived neuronal networks coupled to MEAs during long-term development},
journal = {Journal of Neural Engineering}
}

@article{habibey2022longterm,
  title={Long-term morphological and functional dynamics of human stem cell-derived neuronal networks on high-density micro-electrode arrays},
  author={Habibey, Rouhollah and Striebel, Johannes and Schmieder, Felix and Czarske, J{\"u}rgen and Busskamp, Volker},
  journal={Frontiers in Neuroscience},
  volume={16},
  pages={951964},
  year={2022},
  doi={10.3389/fnins.2022.951964}
}

@article{WANG2023101967,
title = {Generation of glutamatergic/GABAergic neuronal co-cultures derived from human induced pluripotent stem cells for characterizing E/I balance in vitro},
journal = {STAR Protocols},
volume = {4},
number = {1},
pages = {101967},
year = {2023},
issn = {2666-1667},
author = {Shan Wang and Rick Hesen and Britt Mossink and Nael {Nadif Kasri} and Dirk Schubert}
}

@article{frega2017rapid,
  title={Rapid neuronal differentiation of induced pluripotent stem cells for measuring network activity on micro-electrode arrays},
  author={Frega, Monica and Van Gestel, Sebastianus HC and Linda, Katrin and Van Der Raadt, Jori and Keller, Jason and Van Rhijn, Jon-Ruben and Schubert, Dirk and Albers, Cornelis A and Kasri, Nael Nadif},
  journal={Journal of visualized experiments: JoVE},
  number={119},
  pages={54900},
  year={2017}
}

@article{Chiappalone08,
author = {Chiappalone, Michela and Massobrio, Paolo and Martinoia, Sergio},
title = {Network plasticity in cortical assemblies},
journal = {European Journal of Neuroscience},
volume = {28},
number = {1},
pages = {221-237},
year = {2008}
}

@article{MACCIONE2009241,
title = {A novel algorithm for precise identification of spikes in extracellularly recorded neuronal signals},
journal = {Journal of Neuroscience Methods},
volume = {177},
number = {1},
pages = {241-249},
year = {2009},
issn = {0165-0270},
author = {Alessandro Maccione and Mauro Gandolfo and Paolo Massobrio and Antonio Novellino and Sergio Martinoia and Michela Chiappalone}
}

@ARTICLE{Raykov2022,
  author={Raykov, Yordan P. and Saad, David},
  journal={IEEE Journal of Selected Topics in Quantum Electronics}, 
  title={Principled Machine Learning}, 
  year={2022},
  volume={28},
  number={4: Mach. Learn. in Photon. Commun. and Meas. Syst.},
  pages={1-19},
  }





\end{document}